%% file: Wireless_VMC_R1.tex
\documentclass[journal]{IEEEtran}

\IEEEoverridecommandlockouts

\usepackage{fancyhdr}
\usepackage{epsfig}
\usepackage{threeparttable}
\usepackage{epsf,epsfig}
\usepackage{amsthm}
\usepackage[cmex10]{amsmath}
\usepackage{amssymb, amsfonts}
\usepackage[noadjust]{cite}
\usepackage{dsfont}
\usepackage{subfigure}
\usepackage{color}
\usepackage{soul}
\usepackage{amssymb}
 \usepackage{accents}
  \usepackage{algorithm}
   \usepackage[english]{babel}
    \usepackage{url}
    \usepackage{framed} 
        \usepackage{bm}

\begin{document}
\include{header}
\title{\huge Channel-Driven Monte Carlo Sampling for\\ Bayesian Distributed Learning in Wireless Data Centers}

\author{Dongzhu Liu and Osvaldo Simeone
\thanks{\noindent The authors are with King's Communications, Learning, and Information Processing (KCLIP) lab at the Department of Engineering of Kings College London, UK (emails: dongzhu.liu@kcl.ac.uk, osvaldo.simeone@kcl.ac.uk). The authors have received funding from the European Research Council (ERC) under the European Unions Horizon 2020 Research and Innovation Programme (Grant Agreement No. 725731).}}

\maketitle

\begin{abstract}
Conventional frequentist learning, as assumed by existing federated learning protocols, is limited in its ability to quantify uncertainty, incorporate prior knowledge, guide active learning, and enable continual learning. Bayesian learning provides a principled approach to address all these limitations, at the cost of an increase in computational complexity. This paper studies distributed Bayesian learning in a wireless data center setting encompassing a central server and multiple distributed workers. Prior work on wireless distributed learning has focused exclusively on frequentist learning, and has introduced the idea of leveraging uncoded transmission to enable ``over-the-air'' computing. Unlike frequentist learning, Bayesian learning aims at evaluating approximations or samples from a global posterior distribution in the model parameter space. This work investigates for the first time the design of distributed one-shot, or ``embarrassingly parallel'', Bayesian learning protocols in wireless data centers via consensus Monte Carlo (CMC). Uncoded transmission is introduced not only as a way to implement ``over-the-air'' computing, but also as a mechanism to deploy \emph{channel-driven MC sampling}: Rather than treating channel noise as a nuisance to be mitigated, channel-driven sampling utilizes channel noise as an integral part of the MC sampling process. A simple wireless CMC scheme is first proposed that is asymptotically optimal under Gaussian local posteriors. Then, for arbitrary local posteriors, a variational optimization strategy is introduced.  Simulation results demonstrate that, if properly accounted for, channel noise can indeed contribute to MC sampling and does not necessarily decrease the accuracy level.

\end{abstract}


\section{Introduction}
The availability of massive data sets for machine learning has led to the development of distributed computing platforms that can improve the computational efficiency of training tasks by partitioning a global data set among multiple workers \cite{stoica2017berkeley, li2014scaling,abadi2016tensorflow,dhar2019device,zhang2019deep,celik2018wireless}. Decentralization may come at the cost of a possible degradation in accuracy due to the reliance on local optimization steps \cite{ben2019demystifying}. Moreover, when workers are connected to the central server via capacity-limited channels, communication-related impairments can cause additional performance degradation \cite{zhu2019broadband,amiri2020machine,yang2020federated,lan2020capacity,ren2020scheduling, yang2019scheduling,liu2020privacy, cao2020optimized,zhang2020gradient,celik2018wireless}.


As we will review, 
existing works on wireless distributed learning focus on a standard \emph{frequentist} formulation  of the learning problem, whereby the goal is to identify a single vector of model parameters by minimizing (a function of) the training loss  \cite{amiri2020machine, liu2020privacy, cao2020optimized, zhu2020one, Xing2020Federated}. Much research has been specifically carried out on studying the advantages of \emph{uncoded transmission} to enable ``over-the-air'' computing \cite{amiri2020machine, zhu2019broadband, liu2020privacy, cao2020optimized}. Frequentist learning is known to be limited in its ability to quantify epistemic uncertainty, i.e., uncertainty in the model parameter space, yielding overconfident decisions \cite{guo2017calibration,lakshminarayanan2016simple}.  In contrast, \emph{Bayesian} learning provides a principled way to obtain predictive models that account for epistemic uncertainty. This is done by optimizing not over individual model parameter vectors, but over distributions in the model parameter space \cite{barber2012bayesian,simeone2017brief}. Other advantages of Bayesian learning over frequentist learning include the capacity to incorporate prior knowledge, to guide active learning, and to enable continual learning \cite{barber2012bayesian}. Practical implementations of Bayesian learning involve either approximating the posterior distribution over the model parameters -- an approach known as \emph{variational inference (VI)}, or producing approximate samples from the posterior -- an approach known as \emph{Monte Carlo (MC) sampling} \cite{angelino2016patterns}.

This paper considers, for the first time, distributed Bayesian learning in wireless networks. We specifically focus on a wireless data center setting as illustrated in Fig. \ref{Fig: sys} \cite{celik2018wireless}, and on the standard class of ``embarrassingly parallel'' \emph{consensus Monte Carlo (CMC)} protocols \cite{scott2016bayes, rabinovich2015variational}. In wireless data centers, the global data set is available at the central server, which distributes subsets of data points to the workers. Wireless data centers have been introduced as an alternative to conventional wired topologies to reduce cost and to cope with unbalanced traffic \cite{celik2018wireless}. In CMC protocols, first, each worker produces samples from a local posterior based on the locally available data; then, the server carries out a one-shot aggregation of the received messages in order to approximate samples from the global posterior distribution. 

As a key contribution, we introduce the novel idea of \emph{channel-driven MC sampling}. We specifically demonstrate that, under uncoded wireless transmission, channel noise can be utilized as an integral component of MC sampling. As a result, if properly accounted for, channel noise may not cause harm to the performance of MC-based distributed Bayesian learning. This is in stark contrast to wireless distributed learning based on frequentist principles, for which channel noise is generally detrimental to the learning performance. {\color{black}An exception to this statement was demonstrated in \cite{sery2021over}, which argues that the presence of noise can be useful when training using non-convex objectives.}


\subsection{Wireless Distributed Learning}
Most work on wireless distributed learning focuses on the ``federated'' setting, in which data originates at the workers and it remains unknown to the server. In contrast, in a wireless data center system, the server has access to the global data set, and it distributes data to the workers to benefit from computational parallelism. In both cases, distributed learning protocols can be embarrassingly parallel, that is, one-shot, or iterative. In the former case, which we focus on, the workers carry out local computations and communicate only once with the server, which aggregates the results of the local computations \cite{neiswanger2013asymptotically, scott2016bayes,rabinovich2015variational}; while, in the latter, local computing and communication steps are alternated. To counteract hostile channel fading and additive channel noise conditions in wireless distributed learning, various strategies have been proposed including optimal power control \cite{liu2020privacy, cao2020optimized,zhang2020gradient} and over-the-air computing \cite{zhu2019broadband, amiri2020machine, yang2020federated}.  Over-the-Air computing utilizes non-orthogonal multi-access (NOMA) and uncoded transmission to enable the aggregation of signals transmitted by the workers ``over-the-air", leveraging the waveform-superposition property of a multi-access channel. For the wireless data center setting, existing works address issues such as optimizing the number of workers to balance the tradeoff between communication latency and computation time \cite{song2020wireless}, and designing data partition strategy to cope with straggling workers \cite{lyu2019optimal}.

\subsection{Distributed Bayesian Learning} 

To the best of our knowledge, as reviewed above, prior work on distributed wireless learning -- in both federated and wireless data center settings -- is based on frequentist learning. As mentioned, since exact Bayesian learning is generally intractable, existing Bayesian learning schemes are based on approximate methods leveraging either VI or MC sampling. VI assumes a parametric family for the posterior distribution of the model parameters, and its performance is generally limited by the bias due to the variational approximation \cite{angelino2016patterns}. In contrast, MC sampling aims at obtaining samples from the posterior distribution, and is asymptotically exact in the number of samples. Given their reliance on parametric optimization, distributed VI schemes can be implemented in a manner similar to existing frequentist protocols in federated learning, including local stochastic gradient descent (SGD) optimization and global aggregation \cite{corinzia2019variational,kassab2020federated}. 

Distributed MC sampling can also be implemented either in a one-shot fashion or an iterative manner. Most common is the former class of protocols, which are known as CMC \cite{scott2016bayes}. The key problem in the design of CMC schemes is how to aggregate the local samples provided by the workers \cite{neiswanger2013asymptotically}. In this regard, by approximating the posterior distribution as Gaussian, one can obtain optimal weights in closed form, which can be estimated using the local samples \cite{scott2016bayes}. For general distributions, reference \cite{rabinovich2015variational} proposed to design the aggregation function via a VI formulation. Less well studied are iterative protocols, which include distributed stochastic gradient Markov Chain MC strategies \cite{ahn2014distributed}. These protocols are inherently sequential, making parallel implementations challenging.

All prior works on CMC assume ideal communication. 
Nevertheless, the existence of  channel noise in wireless data centers can severely distort the distribution of the received samples. This observation motivates us to seek optimized aggregation strategies that account for the impairments caused by wireless transmission. We refer to the resulting setting as \emph{wireless CMC}. 

%

\subsection{Contributions, Organization, and Notations}

\begin{figure}[t]
\centering
\includegraphics[width=8 cm]{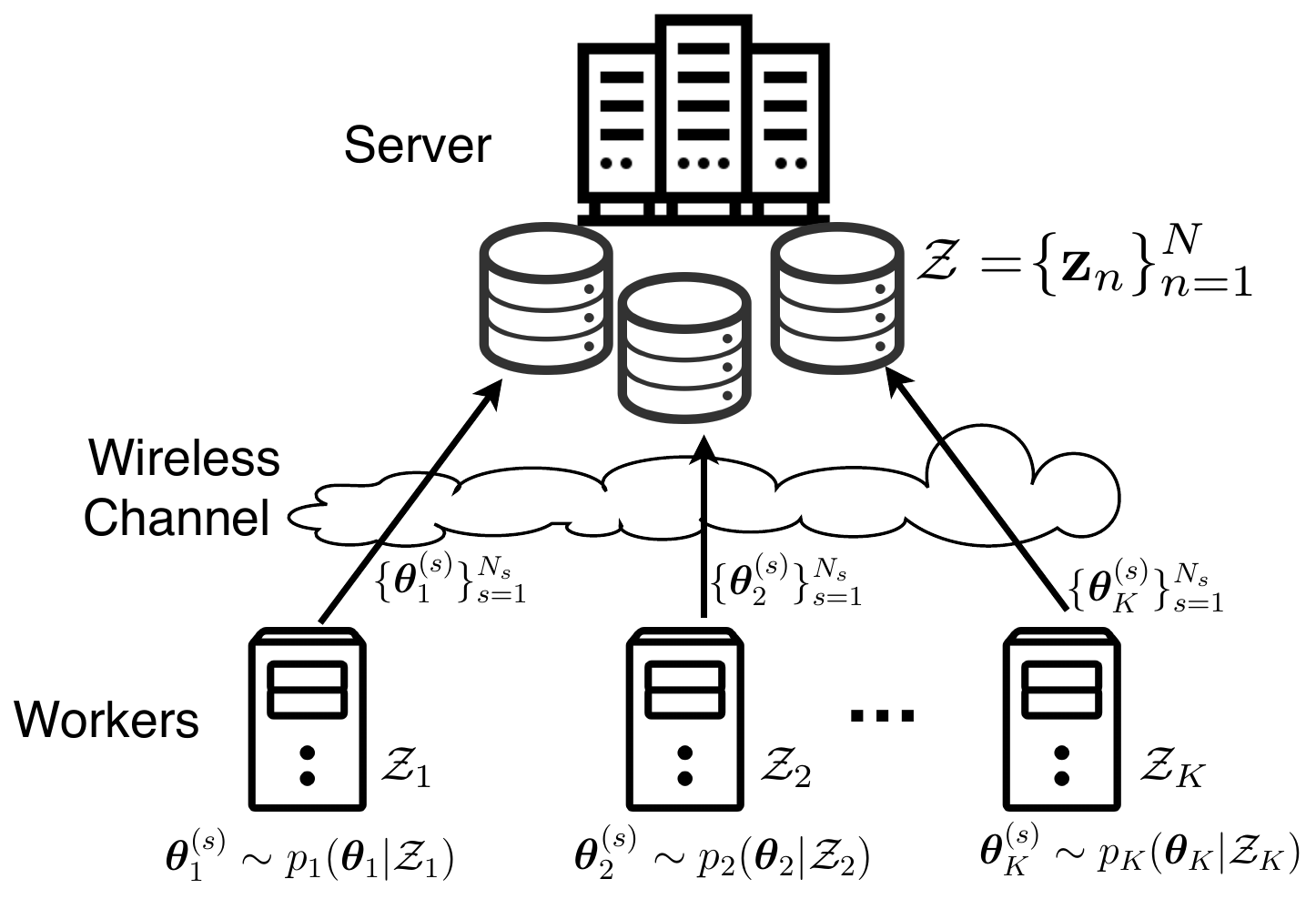}
\caption{Bayesian learning via consensus Monte Carlo (CMC), i.e., embarrassingly parallel distributed MC sampling, in a wireless data center.}
\label{Fig: sys}
\end{figure}

In this paper, we study for the first time the design of wireless CMC protocols for distributed one-shot Bayesian learning in a wireless data center setting. As illustrated in Fig. \ref{Fig: sys}, the global data set is partitioned into disjoint subsets for the distributed workers to enable the implementation of CMC. After local sampling, each worker uploads the produced local samples  to the server via uncoded transmission via either orthogonal or non-orthogonal multi-accesses protocols.  At the server, the received signals are aggregated by using an optimized function to produce approximated samples from the global posterior. The main findings and contributions of the paper can be summarized as follows.


\noindent $\bullet$ {\bf Wireless Gaussian Consensus Monte Carlo (WGCMC): } To start, we study as in \cite{scott2016bayes} the baseline case in which the local posteriors follow Gaussian distributions, and we introduce a wireless CMC protocol that is asymptotically optimal with respect to the number of produced samples.  The solution automatically adapts to the signal-to-noise ratio (SNR) level, and, under Gaussian local posteriors, it is demonstrated to leverage channel noise as a component of the sample generation process,  instead of a nuisance to be mitigated. We consider orthogonal multiple access (OMA), as well as NOMA  with over-the-air computing to enhance spectral efficiency.  

\noindent $\bullet$ {\bf Wireless Variational Consensus Monte Carlo (WVCMC):}  We then target  the general case with arbitrary  local posteriors. We propose a variational design methodology that aims at reducing the Kullback-Liebler (KL) divergence of the distribution of the aggregated samples with respect to the true global posterior.  The proposed principled solution approach is based on SGD steps at the server. 
 
\noindent $\bullet$ {\bf Experiments:} 
 We provide extensive numerical results on both synthetic and real data sets to demonstrate the advantages of channel-driven CMC.

{\bf Organization:}  The remainder of the paper is organized as follows. 
Section~\ref{sec: system model} introduces the system model. 
Section~\ref{sec: GCMC} introduces the necessary background on CMC. 
Section~\ref{sec: tx rx} presents the general form of signal design for wireless CMC, while WGCMC is introduced in Section \ref{sec: WGCMC} and WVCMC in Section~\ref{sec: WVCMC}.  
Section~\ref{sec: sim} provides numerical results, followed by conclusions in Section~\ref{sec: conclusions}.

{\bf Notations:} The following notations are used throughout the paper.  $\otimes$ denotes Kronecker product. ${\bf 1}_l$ is all-ones $l \times 1$ vector. $\delta[\cdot]$ is the Dirac delta function. $\bA^{-1/2}=\bU {\bm \Lambda}^{-1/2} \bU^{\sf T}$ is the operation for any positive semidefinite matrix $\bA=\bU {\bm \Lambda} \bU^{\sf T}$, where $\bU$ is comprised by singular vectors and   $ {\bm \Lambda} $ represents the non-zero diagonalized singular values.  $[\bA]^+$ sets the negative singular values of $\bA$ to 0, and ${\mathcal{I}}(\cdot)$ is the indicator function.

\section{System Model} \label{sec: system model}

\subsection{Setting}
As shown in Fig. \ref{Fig: sys}, we consider a wireless data center setting \cite{celik2018wireless} including a server and $K$ workers connected to it via a shared wireless channel.  The global dataset $\mathcal{Z}$ at the server consists of $N$ data samples $\mathcal{Z}= \{\bz_n\}_{n=1}^N$.  For supervised learning applications, each data sample $\bz_n=(\bu_n, v_n)$ is in turn partitioned into a covariate vector $\bu_n$ and a label $v_n$; while, for unsupervised learning applications such as generative modeling, it consists of a single vector $\bz_n$ (see, e.g., \cite{simeone2017brief}). The goal of the system is to carry out Bayesian learning via Monte Carlo (MC) sampling. This entails generating samples that are (approximately) distributed according to posterior distribution $p({\bm \theta} | \mathcal{Z} )$ of model parameter ${\bm \theta} \in \mathbb{R}^d$ given data set $\mathcal{Z}$.  
Samples from the posterior can be used to obtain ensemble predictive distributions that approximate the optimal Bayesian predictor (see, e.g., \cite{barber2012bayesian,murphy2012machine, simeone2017brief}). 

{\color{black}Denote as $p(\bz | {\bm \theta} )$ the likelihood of model parameter vector $\bm \theta$ for data point $\bz$, and as $p({\bm \theta})$  the prior distribution of the model parameter vector.  Under the standard assumption of conditionally independent and identical (i.i.d.) data points, the global posterior distribution can be written as 
\begin{align}\label{eq: global posterior}
\text{(Global posterior)}  \quad p({\bm \theta} | \mathcal{Z} ) \propto p({\bm \theta}) \prod_{n=1}^N p(\bz_n | {\bm \theta} ). 
\end{align}}
 Note that, for supervised learning problems, the likelihood can be written as $p(\bz|{\bm \theta})=p(\bu,v|{\bm \theta})$ for generative models and as $p(\bz|{\bm \theta})=p(v|\bu,\bm \theta)$ for discriminative models \cite{simeone2017brief}. 
We make the key assumption that exact sampling from the posterior $p({\bm \theta} | \mathcal{Z} )$ is computationally intensive due to the size of the data set $\mathcal{Z}$. Possible solutions include  adaptive subsampling procedures \cite{bardenet2014adaptive} and efficient stochastic gradient-based algorithms such as stochastic gradient Langevin dynamics (SGLD) \cite{welling2011bayesian}, which require limited access to data at the server at any given time. This paper considers an alternative, and potentially more effective approach that leverages the presence of the $K$ distributed workers in the wireless data center system in Fig.~\ref{Fig: sys}.

We specifically propose, and study, the wireless implementation of distributed MC sampling solutions,  whereby workers perform parallel MC sampling on distinct subsets of the data \cite{angelino2016patterns}. 
We focus on the class of embarrassingly parallel strategies, known as consensus Monte Carlo (CMC), in which communication from workers to server occurs only once, after local processing at the workers is completed.  
In CMC, the global dataset $\mathcal{Z}$ is partitioned into $K$ non-overlapping subsets $\mathcal{Z}_1, \cdots, \mathcal{Z}_K$, each of size $N_k$, and each data set $\mathcal{Z}_k$ is allocated to the $k$-th worker. We have $\sum_{k=1}^K N_k = N$.  {\color{black}Data delivery is offline, and hence one can consider data to be available without distortion at the workers.}
 It is assumed that the $k$-th worker has sufficient computational resources to generate samples from the subposterior $p_k({\bm \theta}|\mathcal{Z}_k)$. The subposterior depends on the allocated subset $\mathcal{Z}_k$ and on the underweighted prior $p({\bm \theta}) ^{1/K}$ as 
\begin{align}\label{eq: subposterior}
\text{(Subposterior)} \quad p_k({\bm \theta}|\mathcal{Z}_k) \propto   p({\bm \theta}) ^{1/K} p(\mathcal{Z}_k|{\bm \theta}) ,
\end{align}
where we have defined the local likelihood function as $p(\mathcal{Z}_k|{\bm \theta})=\prod_{\bz\in\mathcal{Z}_k}p(\bz|{\bm\theta})$. {\color{black}This assumption is justified by the fact that the complexity of MC sampling scales up with the size of the underlying data set \cite{scott2016bayes, rabinovich2015variational, angelino2016patterns} (see Appendix \ref{app: sampling complexity} for some additional discussion on this point).}
Note that the global posterior \eqref{eq: global posterior} can be expressed as  the product of subposteriors \eqref{eq: subposterior}, upon normalization,~as  
\begin{align}
 p({\bm \theta}|\mathcal{Z}) \propto    \prod_{k=1}^Kp_k({\bm \theta}|\mathcal{Z}_k).  \label{eq: fact global posterior}
\end{align}

Each worker $k=1,\dots, K$ generates $S$ samples ${\bm \theta}_k^{(s)}\sim  p_k({\bm \theta}|\mathcal{Z}_k)$ for $s=1,\dots,S$ and transmits them one-by-one, as detailed in the next section, over the wireless channel to the server. Based on the received signals on the wireless channel, the server aims at obtaining samples ${\bm \theta}^{(s)}$ for $s=1,\dots, S$ that are approximately drawn from the global posterior $p({\bm \theta}|\mathcal{Z})$ by applying an aggregation function $F(\cdot)$. We note that the aggregation function $F(\cdot)$ may depend on the data set $\cZ$, which is available at the server. 

In existing CMC methods \cite{scott2016bayes, rabinovich2015variational}, the samples $\{{\bm \theta}_k^{(s)}\}_{k=1}^K$ are assumed to be received noiselessly to the server, and the aggregation function is chosen as the linear combination  
\begin{align}
{\bm \theta}^{(s)}=F({\bm \theta}_1^{(s)},\cdots,{\bm \theta}_K^{(s)})=\sum_{k=1}^K \bW_k {\bm \theta}_k^{(s)} \label{eq: agg func}
\end{align}
for some $d \times d$ weight matrices $\bW_1,\cdots, \bW_k$. {\color{black}CMC does not require  feedback from the edge server or other communications beyond the transmission of the local posterior samples from workers to the server.}  The weight matrices are optimized based on the samples $\{\{{\bm \theta}^{(s)}_k\}_{k=1}^K\}_{s=1}^S$ and on data set $\cZ$ \cite{scott2016bayes, rabinovich2015variational}. CMC is successful when the distribution $q_F({\bm \theta})$ of the samples ${\bm \theta}^{(s)}$ computed at the server is a good approximation of the global posterior $p({\bm \theta}|\mathcal{Z})$.

{\color{black}In this paper, we will consider for the first time the case in which the samples $\{\{{\bm \theta}_k^{(s)}\}_{k=1}^K\}_{s=1}^S$  have to be communicated to the server over a wireless channel, and demonstrate the novel use of channel noise for Monte Carlo sampling. The proposed channel-driven sampling is enabled by the use of analog communication for the transmission of the samples $\{{\bm \theta}^{(s)}_k\}_{k=1}^K$, whereby channel noise is directly added to the samples produced and transmitted by the workers. The CMC aggregation function \eqref{eq: agg func} can be implemented using both orthogonal and non-orthogonal multiple access schemes, which have distinct advantages and disadvantages.  With orthogonal multiple access (OMA) transmission, the local samples $\{{\bm \theta}^{(s)}_k\}_{k=1}^K$ are received separately. Therefore, channel noise acts as a distinct  source of randomness for each sample, and arbitrary weight matrices $\{\bW_k\}_{k=1}^K$ can be applied at the server. In contrast, under non-orthogonal multiple access (NOMA) transmission, AirComp can increase communication efficiency, but at the costs of sharing the channel noise across samples and of limiting the choice of the weight matrices $\{\bW_k\}_{k=1}^K$ (see Sec. \ref{sec: tx rx}).
}

%
%

\subsection{Communication Model}\label{sec: comm model}
Following recent work on communication for machine learning \cite{amiri2020blind, zhu2020one, yang2020federated, zhu2019broadband,  zhu2020toward}, we assume that workers communicate with the server via a  general vector wireless communication model that may account for multi-carrier and/or single/multi-antenna setups. Accordingly, each worker transmits $m_t$  symbols, over time, space, and/or frequency domains, and the server receives $m_r$ symbols over a channel described by a real matrix and additive noise.  We focus on a real channel model with no loss of generality, since complex vector channels can be described in terms of real vector channel of twice the size of the original model \cite{liu2020privacy}. We will specifically consider two general classes of communication protocols, namely orthogonal multiple access (OMA) and non-orthogonal multiple access (NOMA).

\subsubsection{Orthogonal multiple access (OMA)} Under OMA,  workers transmit over orthogonal communication resources, and the $m_r \times 1$ received signal from the $k$-th worker is 
\begin{align} \label{eq: rev signal}
\by_k=\bH_k\bx_k+\bn_k,
\end{align}
where $\bH_k\in \mathbb{R}^{m_r\times m_t}$ is the channel matrix between worker $k$ and server; $\bx_k\in \mathbb{R}^{m_t}$ is the signal transmitted by worker $k$; and $\bn_k\in\mathbb{R}^{m_r}$ is channel noise, with entries being independent and identically distributed (i.i.d.) $\mathcal{N}(0, N_0)$ random variables (rvs).  Specifically, in each communication block, a given worker $k$ is scheduled to transmit a sample ${\bm \theta}_k^{(s)}$. We focus on analog, i.e., uncoded transmission strategies, whereby ${\bm \theta}_k^{(s)}$ is mapped through a linear transformation to $\bx_k$, as we will detail in Sec. \ref{sec: tx rx}. 

\subsubsection{Non-orthogonal multiple access (NOMA)} Under NOMA, all workers transmit simultaneously, and the $m_r \times 1$ received signal given by the noisy superposition
\begin{align} \label{eq: rev signal noma}
\by=\sum_{k=1}^K\bH_k\bx_k+\bn,
\end{align}
where $\bH_k\in \mathbb{R}^{m_r\times m_t}$ is the channel matrix between worker $k$ and server; $\bx_k\in \mathbb{R}^{m_t}$ is the signal transmitted by worker $k$; and $\bn\in\mathbb{R}^{m_r}$ is channel noise, with entries being i.i.d. $\mathcal{N}(0, N_0)$ random rvs. As for OMA, each signal $\bx_k$ encodes a sample ${\bm \theta}_k^{(s)}$ produced by worker $k$ through a linear transformation.

In both cases, the channel matrices generally change from one communication block, given by \eqref{eq: rev signal} or \eqref{eq: rev signal noma}, to another. {\color{black}The channel matrix $\bH_k$ and its statistics are assumed to be known at  worker $k$. This can be achieved via the broadcasting of pilots by the server in time-division duplex (TDD) systems, or via uplink pilot transmission and feedback from the server for  frequency-division duplex (FDD) systems.}  An important aspect of the model is that, in order to receive information from all devices, NOMA can operate with a fraction $1/K$ of the resources of OMA, since the server obtains a superposition of all $K$ signals in a single communication block. This property  has been successfully leveraged by AirComp strategies based on analog transmission \cite{zhu2018mimo, amiri2020machine, yang2020federated, zhu2020over, liu2020privacy}.

For both OMA and NOMA, the transmit per-block power constraint is set as the inequality
\begin{align}\label{eq: power constraint}
\E[\|\bx_k\|^2]\leq P,
\end{align}
where the expectation is taken with respect to the distribution of the channels and of the encoded samples. The average over the channel implies consideration of a long-term power constraint. 
{\color{black}This is sensible for the common setting in which the workers transmit multiple samples $S>1$ in order to enable posterior averaging at the server, e.g., for ensemble prediction (see Sec. \ref{sec: sim}). }
A short-term power constraint can be accounted for in a similar way. 

Throughout the paper, we assume the conditions $m_t\geq m_r\geq d$, where we recall that $d$ is the dimension of the model vector $\bm \theta$. {\color{black} As is common in the Bayesian learning literature, the value of $d$ is generally small in prior CMC works \cite{rabinovich2015variational,scott2016bayes}. For larger models, i.e., for larger $d$, the transmission of one sample would require multiple coherence blocks and/or use sufficiently large antenna arrays. This setting can be modelled by introducing block diagonal channel matrices, including the channel matrices for each block on the block diagonal.} The integration of analog compression techniques (see, e.g., \cite{abdi2020analog, abdi2020quantized}) with the wireless CMC strategies to be developed here can alleviate this problem and is left for future work. 
Finally, we assume that the channel matrices $\{\bH_k\}_{k=1}^K$ have full rank with probability one. This is a common assumption in many studies on multi-antenna wireless communication (see, e.g., \cite{jafar2011interference}). 

%
%

\section{Background: \\ Gaussian Consensus Monte Carlo}\label{sec: GCMC}
In this section, we review the Gaussian CMC (GCMC) scheme introduced in \cite{scott2016bayes} that assumes noiseless, ideal communication. Accordingly, GCMC assumes that the local samples $\{{\bm \theta_k^{(s)}}\}_{k=1}^K$, for $s=1,\dots,S$ are available at server. We will argue in Sec. \ref{sec: WGCMC} that a direct application of this scheme in the wireless setting under study is strictly suboptimal, even in the asymptotic regime of large number of samples, $S$, motivating the introduction of alternative schemes. 



GCMC is based on the observation that, if each local subposterior $p_k({\bm \theta}|\mathcal{Z}_k)$ is a  Gaussian $\cN({\bm \mu}_k,\bC_k)$ distribution with invertible covariance matrix $\bC_k$, the global posterior is the Gaussian $\cN({\bm \mu},\bC)$ distribution, where ${\bm \mu}=\bC (\sum_{k=1}^K \bC_k^{-1}{\bm \mu}_k) $ and $\bC=(\sum_{k=1}^K\bC_k^{-1})^{-1}$ . If follows that, given local samples ${\bm \theta}_k \sim \cN({\bm \mu}_k,\bC_k)$, for $k=1,\dots,  K$, a global sample ${\bm \theta}\sim \cN({\bm \mu},\bC)$ can be obtained by applying the linear combination \eqref{eq: agg func} with weight matrices 
\begin{align}
\bW_k=\Big(\sum_{k=1}^K\bC_k^{-1}\Big)^{-1} \bC_k^{-1}. \label{eq: w GCMC}
\end{align}

Since the matrices $\bC_1,\cdots, \bC_K$ are not known, GCMC estimates them by using the $S$ samples $\{{\bm \theta}_k^{(s)}\}_{k=1}^K$ with $s=1,\dots,S$ received from the workers. In particular, given the $S$ samples $\{{\bm \theta}_k^{(s)}\}_{s=1}^S$ received form worker $k$, covariance $\bC_k$ is  estimated  as \begin{align}\label{eq: GCMC}
\widehat{\bC}_k = \frac{1}{S-1} \sum_{s=1}^{S} \l({\bm \theta}_k^{(s)} - \widehat{\bm \mu}_k \r) \l({\bm \theta}_k^{(s)} - \widehat{\bm \mu}_k \r)^{\sf T}, 
\end{align}
with $\widehat{\bm \mu}_k= \frac{1}{S} \sum_{s=1}^{S} {\bm \theta}_k^{(s)} $. Then, the function \eqref{eq: agg func} is applied using the estimate $\widehat{\bC}_k$ \eqref{eq: GCMC} in lieu of $\bC_k$ in the weights~\eqref{eq: w GCMC}. Accordingly, GCMC computes each global sample ${\bm \theta}^{(s)}$ as 
\begin{align}\label{eq: est GCMC result}
\text{(GCMC)} \quad {\bm \theta}^{(s)}= \sum_{k=1}^K  \underbrace{  \Big(  \sum_{k'=1}^K\widehat{\bC}_{k'}^{-1}\Big)^{-1} \widehat{\bC}_k^{-1}} _{\bW_k} {\bm \theta}_k^{(s)} . 
\end{align}
Note that, as a result, the function \eqref{eq: est GCMC result} applied by the server is non-linear in the samples $\{\{{\bm \theta}^{(s)}\}_{k=1}^K\}_{s=1}^S$.  It is also noted that GCMC does not require the use of the data set $\cZ$ at the server. {\color{black}The matrix inversion in \eqref{eq: est GCMC result} incurs prohibitive computational complexity for high-dimensional models. One possible solution, as proposed in \cite{rabinovich2015variational,scott2016bayes}, is to restrict the weight matrices $\bW_k$ to be diagonal, with each element corresponding to the inverse of the marginal variance of the samples.} 

{\color{black}While GCMC is defined under the assumption of Gaussian subposteriors, reference \cite{scott2016bayes} argues that the choice \eqref{eq: w GCMC}--\eqref{eq: GCMC} for the combining matrices $\{\bW_k\}_{k=1}^K$ is suitable for other settings as well. This is theoretically well justified for large data sets, i.e., when the data set sizes $\{N_k\}_{k=1}^K$ are large as compared to the model dimension, since the subposteriors in this case can be well approximated by Gaussian distributions according to Bernstein-von Mises theorem \cite{johnstone2010high}. In this paper, we will adopt throughout a linear aggregation as in \eqref{eq: agg func}, although the methods to be introduced in the next sections apply to more general parametric aggregation functions.}

\section{Wireless Consensus Monte Carlo: \\Signal Design}\label{sec: tx rx}
Unlike prior works on CMC \cite{rabinovich2015variational,scott2016bayes}, including the baseline GCMC scheme reviewed in the previous section, in the following sections, we propose solutions that account for the  wireless communication channel between workers and server. 
 A key novel idea introduced by this work is that the noise added by the wireless channel 
 may not cause any performance degradation as compared to an ideal scenario with noiseless transmission: Noise can contribute to MC sampling, and is not necessarily a nuisance to be mitigated. In this section,  we describe the general form of the considered uncoded transmission strategies for CMC. The next two sections propose specific wireless CMC protocols. Throughout, we focus on analog, or uncoded, transmission strategies with the aims of leveraging channel noise for sampling and enabling AirComp solutions via NOMA. Comparison with digital schemes follow along similar lines as in \cite{amiri2020machine, Xing2020Federated}, and is left as future work. 
 
%

\subsection{Signal Design in OMA} 
In OMA, in each communication block,  a single sample ${\bm \theta}_k^{(s)} $ is transmitted by the scheduled device $k$ to the server. Dropping temporarily the superscript $s$ to simplify the notation, we assume uncoded transmission of the form 
\begin{align}\label{eq: tx signal}
\text{(Transmitted Signal)}\quad \bx_k&= \bU_k  {\bm \theta}_k,
\end{align}
where $\bU_k$ is an $\mathbb{R}^{m_t\times d}$ precoding matrix. We assume that $\bU_k$ factorizes into a zero-forcing pre-equalization matrix $\bH_k^\dagger=\bH_k^{\sf T}(\bH_k \bH_k^{\sf T} )^{-1}$ and a $m_r\times d$ encoding matrix as $\bU_k=\bH_k^\dagger \bE_k$. The use of  zero-forcing  is aligned with the majority of studies on uncoded transmission for machine learning \cite{zhu2018mimo,zhu2020over,wen2019reduced,liu2020privacy}, although generalizations are possible. The encoding matrix $\bE_k$ is fixed, and not subject to optimization. For instance, if $m_r$ is a multiple of $d$, e.g., $m_r=ld$ for some integer $l$, one can set $\bE_k\propto {\bf 1}_l \otimes \bI_d$  to implement a form of analog repetition coding. Optimization of analog coding strategies  is left for future work. The encoding matrix $\bE_k$ is subject to the power constraint \eqref{eq: power constraint}, which imposes the inequality 
\begin{align}
\E[\|\bx_k\|^2]=\E[\tr(\bU_k {\bm  \theta}_k {\bm \theta}_k^{\sf T}\bU_k ^{\sf T})]\leq P. \label{eq: origin pc}
\end{align}
Since the distribution of the samples $\{{\bm \theta}_k\}_{k=1}^K$ is unknown, we will specifically impose \eqref{eq: origin pc} by averaging over the available $S$ samples as 
\begin{align}
\frac{1} {S}\sum_{s=1}^S\tr\l( \E\big[ \big(\bH_k \bH_k^{\sf T})^{-1}\big]\bE_k {\bm \theta}_k^{(s)}  ({\bm \theta}_k^{(s)})^{\sf T} \bE_k^{\sf T}  \r) \leq P . \label{eq: approx pc}
\end{align}
The inequality \eqref{eq: approx pc} can be thought of an energy constraint across the transmission of the $S$ samples for each worker $k$. 
For the example of analog repetition coding, constraint \eqref{eq: approx pc} is satisfied with $\bE_k=\sqrt{{P_k}} {\bf 1}_l \otimes \bI_d $,  where $P_k= PS/\sum_{s=1}^S\tr\big( \E\big[ \big(\bH_k \bH_k^{\sf T})^{-1}\big]({\bf 1}_l \otimes \bI_d ) {\bm \theta}_k^{(s)}  ({\bm \theta}_k^{(s)})^{\sf T} ({\bf 1}_l \otimes \bI_d )^{\sf T}  \big) $. 

 At the receiver side, for OMA, we propose to aggregate the received signals, $\by_1,\cdots, \by_K$ in a manner akin to  \eqref{eq: agg func} as 
\begin{align}\label{eq: est theta}
\text{(Aggregation OMA)}\quad {\bm \theta}=F(\by_1,\cdots,\by_K)=\sum_{k=1}^K \bW_k\by_k,
\end{align}
where $\bW_1, \cdots, \bW _K$ are  $d\times m_r$ weight matrices.  Note that the aggregation in \eqref{eq: est theta} is applied separately for each set of received signals $\{{\by}_k^{(s)}\}_{k=1}^K$. The weights $\{\bW_k\}_{k=1}^K$ are to be optimized based on the received signals $\{\by_k^{(s)}\}_{k=1}^K$ for all samples $s=1,\dots,S$, as well as, possibly, based on the data set $\cZ$.

%

\subsection{Signal Design in NOMA}   
Under NOMA, in each communication block, all workers transmit one of their local samples simultaneously by using uncoded transmission as in \eqref{eq: tx signal}. As in OMA, matrix $\bU_k$ is factorized as the product $\bU_k=\bH_k^\dagger \bE_k$. Furthermore, to enable the implementation of AirComp in a manner similar to \cite{zhu2018mimo,wen2019reduced,liu2020privacy}, we assume that all the transmitted signals are aligned by using the same encoding matrix, which denoted as $\bE_k=\bE$ for all $k=1,\dots, K$. This matrix is subject to power constraint  \eqref{eq: power constraint} for all workers $k=1,\dots, K$ as 
\begin{align} \label{eq: power constraint NOMA}
\frac{1} {S}\sum_{s=1}^S\tr\l( \E\big[ \big(\bH_k \bH_k^{\sf T})^{-1}\big]\bE {\bm \theta}_k^{(s)}  ({\bm \theta}_k^{(s)})^{\sf T} \bE^{\sf T}  \r) \leq P . 
\end{align}
 For the mentioned example of  repetition coding, the encoding matrix should be set as $\bE=\sqrt{{\min_kP_k}} {\bf 1}_l \otimes \bI_d $ in order to satisfy the constraint in \eqref{eq: power constraint NOMA}, where the value of $P_k $ is same as for~OMA. 

The received signal \eqref{eq: rev signal noma} in NOMA carries out aggregation ``on air". Applying a single weight matrix $\bW\in\mathbb{R}^{d\times m_r}$ yields the global sample
\begin{align}
\text{(Aggregation NOMA)} \quad {\bm \theta}=F(\by) = \bW \by.  \label{eq: est theta NOMA}
\end{align}
The weight matrix $\bW$ is again applied separately for each received signal $\by^{(s)}$, and is optimized by using all received signals $\{\by^{(s)}\}_{s=1}^S$, as well as, possibly, data set $\cZ$. {\color{black}Note that, while in OMA one can potentially apply any aggregation function to the received samples $\{\{\by_k^{(s)}\}_{s=1}^S\}_{k=1}^K$, over-the-air computing via NOMA limits the aggregation to the linear superposition \eqref{eq: rev signal noma}
 carried out by the channel.}

To unify notation, we will write $\bW=\{\bW_k\}_{k=1}^K$ for the collection of weight matrices for OMA.

\section{Wireless Gaussian Consensus Monte Carlo}\label{sec: WGCMC}

In this section, we present a first wireless CMC protocol that is inspired by GCMC, and referred to as wireless Gaussian consensus Monte Carlo (WGCMC). Unlike GCMC, WGCMC can be proved to be asymptotically exact with respect to  the number of samples, $S$, under the described wireless channel model  when the  subposteriors are Gaussian. The limitations of WGCMC will motivate the introduction of a general optimization procedure for the decoding matrices in the next section. 

To proceed, as in GCMC (see Sec. \ref{sec: GCMC}), we assume that each local subposterior $p_k({\bm \theta}|\mathcal{Z}_k)$ is a  Gaussian $\cN( {\bf 0} ,\bC_k)$ distributions,  so that the global posterior is the Gaussian $\cN({\bf 0} ,\bC)$ distribution, where $\bC=(\sum_{k=1}^K\bC_k^{-1})^{-1}$. Furthermore, in order to simplify the notation, we set $m_t=m_r=d$, and we choose the trivial encoding matrices $\bE_k=\sqrt{{P_k}}\bI_d $ for OMA and  $\bE=\sqrt{\min_k{P_k}}\bI_d $ for NOMA, where  $P_k= {PS}/\sum_{s=1}^S\tr\big( \E\big[ \big(\bH_k \bH_k^{\sf T})^{-1}\big]{\bm \theta}_k^{(s)}  ({\bm \theta}_k^{(s)})^{\sf T}   \big) $. The generalization to any set of encoding matrices follows directly from the discussion below. 

\subsection{Wireless GCMC: OMA}
Let us consider OMA first. For any weight matrices $\bW$, the distribution of the output of the aggregated function \eqref{eq: est theta}  is given as 
\begin{align}\label{eq: sample wgcmc oma}
{\bm \theta}=\sum_{k=1}^K{ \bW_k\by_k} \sim \cN\l({\bf 0},\sum_{k=1}^{K} \bW_k\big({P_k}\bC_k+{N_0} \bI\big) \bW_k^{\sf T}  \r). 
\end{align}
Accordingly,  if we know the covariance matrices $\{\bC_k\}_{k=1}^K$, the power of encoded sample $P_k$, and noise variance $N_0$, optimal weight matrices can be obtained by solving the following equality 
\begin{align}\label{eq: distr equality}
\sum_{k=1}^{K} \bW_k\big({P_k}\bC_k+{N_0} \bI\big) \bW_k^{\sf T}  =\Big(\sum_{k=1}^K\bC_k^{-1}\Big)^{-1}.
\end{align}
In fact, this condition ensures that the distribution of the global sample ${\bm \theta}=\sum_{k=1}^K \bW_k \by_k $ is exactly the desired posterior, i.e., ${\bm \theta}=\sum_{k=1}^K \bW_k \by_k\sim\cN({\bf 0} ,\bC)$. 
A  solution to this system of equation is given by the matrices 
\begin{align}
\bW_k=\Big(\sum_{k=1}^K\bC_k^{-1}\Big)^{-1} \bC_k^{-1/2}\big({P_k}\bC_k+{N_0} \bI\big)^{-1/2}.  \label{eq: asymp opt}
\end{align}
Solution \eqref{eq: asymp opt} is not computable since the matrices $\bC_1,\cdots, \bC_K$ are unknown. They can be estimated by using the noisy received samples $\{\by_1^{(s)},\cdots, \by_K^{(s)}\}_{s=1}^{S}$ as 
\begin{align}\label{est: ck}
\widehat{\bC}_k= \! \frac{1}{P_k}\l[ \!\frac{1}{ (S-1)}\! \sum_{s=1}^{S} \l(\by_k^{(s)}\!-\! \widehat{\bm \mu}_{\by_k}\! \r)\! \l(\by_k^{(s)} - \widehat{\bm \mu}_{\by_k} \r)^{\sf T} \!\!-\!{N_0} \bI\r]^+,
\end{align}
where $\widehat{\bm \mu}_{\by_k}= \frac{1}{S}\sum_{s=1}^S \by_k^{(s)}$.  Note that the estimate $\widehat{\bC}_k$ in \eqref{est: ck} is positive semidefinite thanks to the use of operation $[\ \cdot \ ]^+$. Plugging \eqref{est: ck} into \eqref{eq: asymp opt} yields WGCMC scheme
\begin{align}\label{eq: est WGCMC result}
&\text{(WGCMC for OMA)} \nn\\
& {\bm \theta}^{(s)}= \sum_{k=1}^K  \underbrace{  \Big(  \sum_{k'=1}^K\widehat{\bC}_{k'}^{-1}\Big)^{-1} \widehat{\bC}_k^{-1/2} (P_k\widehat{\bC}_k+N_0 \bI)^{-1/2}} _{\bW_k} {\by}_k^{(s)} 
\end{align}

The estimated covariance $\widehat{\bC}_k $ in \eqref{est: ck} approaches  the true covariance $\bC_k$ as  $S \rightarrow \infty$. Under this condition, the distribution of the global sample $\bm \theta$ in \eqref{eq: sample wgcmc oma} equals the desired global posterior. In contrast,  the conventional GCMC scheme when applied directly to the received signals, i.e., by replacing scaled received signals $\{\by_k^{(s)}/\sqrt{P_k}\}_{s=1}^S$ for $\{{\bm \theta}_k^{(s)}\}_{s=1}^S$ in \eqref{eq: est GCMC result}, is not asymptotically optimal as $S\rightarrow \infty$ since it disregards the presence of noise on the channel.  


\subsection{Wireless Gaussian CMC: NOMA} \label{sec: WGCMC noma}
To develop an asymptotically optimal solution for NOMA, we further assume that local subposteriors are identical Gaussian $\cN(0, \bC_0)$ distributions, so that the global posterior is the Gaussian $\cN({\bf 0} , \frac{1}{K}\bC_0)$ distribution.   This highly simplified scenario with homogeneous posteriors is considered here to illustrate the potential benefits, but also the possible limitations, of NOMA. 

The distribution of the output of the aggregated function \eqref{eq: est theta}  is given as 
\begin{align}
{ \bW\by} \sim \cN\l({\bf 0},\bW  \big(  \min_k P_k K \bC_0+{N_0} \bI\big) \bW^{\sf T}  \r). 
\end{align}
In a manner similar to OMA, the optimal weight matrix can be obtained by solving the following equality 
\begin{align}\label{eq: distr equality GCMC NOMA}
\bW  \big(  \min_k P_k K \bC_0+{N_0} \bI\big) \bW^{\sf T}  = \frac{1}{K}\bC_0,
\end{align}
and one of solutions is given as 
\begin{align}
\bW=\frac{1}{\sqrt{K}}\bC_0^{1/2}\big(  K \min_k P_k  \bC_0+{N_0} \bI\big) ^{-1/2}. \label{eq: asymp opt NOMA}
\end{align}
The unknown matrix $\bC_0$ can be estimated by using the noisy received samples $\{\by^{(s)}\}_{s=1}^{S}$ as 
\begin{align}\label{est: ck noma}
\widehat{\bC}_0\!= \!\frac{1}{K\min_kP_k}\!\l[ \! \frac{1}{ (S-1)} \! \sum_{s=1}^{S} \! \l( \by^{(s)} \!- \! \widehat{\bm \mu}_{\by} \r) \!\! \l(\by^{(s)}\! - \! \widehat{\bm \mu}_{\by}\!\r)^{\sf T} \!\!-\!{N_0} \bI\r]^+\!\!\!\!,
\end{align}
where $\widehat{\bm \mu}_{\by}= \frac{1}{S}\sum_{s=1}^S \by^{(s)}$. Plugging \eqref{est: ck noma} into \eqref{eq: distr equality GCMC NOMA} yields the WGCMC scheme under NOMA,~i.e., 
\begin{align}\label{eq: est WGCMC result NOMA}
&\text{(WGCMC for NOMA)}\nn\\
&{\bm \theta}^{(s)}= \underbrace{\frac{1}{\sqrt{K}}\widehat{\bC}_0^{1/2}\big(  K \min_k P_k  \widehat{\bC}_0+{N_0} \bI\big) ^{-1/2}} _{\bW} \by^{(s)}. 
\end{align}

As for OMA, the sample ${\bm \theta}=\bW\by$ generated by of WGCMC under NOMA in \eqref{eq: asymp opt NOMA} follows the desired global posterior distribution when  $S \rightarrow \infty$. The potential advantage of NOMA versus OMA is that, with the same number of communication blocks, $T$, the number of received samples is $S=T$ in NOMA while it equals $S=T/K$ in OMA. The large number of samples is generally advantageous for the downstream applications that use the samples, e.g., for ensemble prediction. Furthermore, a larger $S$ leads to a more accurate estimation of the relevant covariance matrices. On the flip side, optimality was only argued here for a simplified homogeneous setting, and one may expect a degraded performance in heterogeneous settings. A strategy that addresses the case of general subposterior is introduced next. 


%
%
%
%

\section{Wireless Variational Consensus Monte Carlo} \label{sec: WVCMC}
The WGCMC schemes proposed in the previous subsection can be applied to any class of subposteriors in a manner similar to GCMC. However, apart form the special case of Gaussian subposteriors, this application is expected to be successful only in the regime of large data sets due to the Bernstein-von Mises Theorem.  In general, the WGCMC scheme is suboptimal. In this section, we introduce an optimization strategy for the design of the weight matrices $\bW$ for OMA and NOMA that is able to adapt to arbitrary subposteriors and channels. 

\subsection{Design Principle}
Inspired by \cite{rabinovich2015variational}, 
we adopt a variational approach to formulate the problem of ensuring that the distribution $q_F({\bm \theta})$ of the aggregated samples \eqref{eq: est theta} for OMA and \eqref{eq: est theta NOMA} for NOMA is close to the true posterior $p({\bm \theta}|\cZ)$.  We recall that reference \cite{rabinovich2015variational} assumed noiseless channels and hence focused on optimizing the aggregation function \eqref{eq: agg func} instead. 
This condition can be expressed as the approximate equality 
\begin{subequations} \label{eq: approx global posterior} 
\begin{align}
 &\text{(Mean Field Approximation for OMA)} \nn\\
 &  p({\bm \theta}|\mathcal{Z}) \approx  q_F({\bm \theta})\nn\\
 &= \!\!\int\!\!\! \bigg( \! \prod_{k=1}^K  p(\by_k)\!\! \bigg) \delta\l[{\bm \theta}  = F(\by_1,\cdots,\by_K)\r] d \by_1, \cdots d \by_K  \label{eq: mf oma y}\\
 &= \!\!\int\!\!\! \bigg( \prod_{k=1}^K p_k({\bm \theta}_k |\mathcal{Z}_k) p(\bn_k) \bigg)\delta\l[{\bm \theta}  = F(\by_1,\cdots,\by_K)\r]    d {\bm \theta}_1, d\bn_1, \nn\\
 & \qquad \qquad  \qquad \qquad \qquad \qquad \quad \qquad \cdots  ,d{\bm \theta}_K, d \bn_K  \!  \label{eq: mf oma theta} \\
 &= \E_{{\bm \theta}_1,\cdots,{\bm \theta}_K, \bn_1,\cdots, \bn_K}\Big[\delta\l[{\bm \theta}  = F(\by_1,\cdots,\by_K)\r] \Big],
  \end{align}
  \end{subequations}
  for OMA, where $p(\bn_k)$ is the $\cN(0,N_0 \bI_{m_r})$ probability density function (pdf) and  $\delta[\cdot]$ is the Dirac delta function; and 
\begin{subequations}\label{eq: approx global posterior NOMA}
\begin{align}
 &\text{(Mean Field Approximation for NOMA)}    \nn\\
 & p({\bm \theta}|\mathcal{Z}) \approx   q_F({\bm \theta})\nn\\
  &=\int  p(\by) \delta\l[{\bm \theta}  = F(\by)\r] d \by \label{eq: mf noma y} \\
&= \int  \bigg(p(\bn) \prod_{k=1}^K p_k({\bm \theta}_k |\mathcal{Z}_k) \bigg) \delta\l[{\bm \theta}  = F(\by)\r] d \bn,  d {\bm \theta}_1,\cdots  ,d{\bm \theta}_K   \qquad \quad \qquad \qquad \label{eq: mf noma theta} \\
&=\E_{{\bm \theta}_1,\cdots,{\bm \theta}_K, \bn} \Big[\delta\l[{\bm \theta}  = F(\by)\r] \Big],      
\end{align}
\end{subequations}
for NOMA, where $p(\bn)$ is the $\cN(0,N_0 \bI_{m_r})$ pdf. The distribution $q_F({\bm \theta})$ in \eqref{eq: mf oma theta} and \eqref{eq: mf noma theta} of the obtained global sample is computed by averaging over the distribution of the local samples $({\bm \theta}_1,\cdots,{\bm \theta}_K)$ and of the channel noise. Note that, for OMA, as per \eqref{eq: mf oma y}, the received signal $\{\by_k\}_{k=1} ^K$ are independent due to the independence of local samples and noise samples.

In order to enforce the approximate equalities \eqref{eq: approx global posterior} through the optimization of the weight matrices $\bW$, we propose to tackle the problem of minimizing the distance of the distributions $q_F({\bm \theta})$ and $  p({\bm \theta}|\mathcal{Z})$ as measured by Kullback-Leibler (KL) divergence 
$${\rm KL}\big(q_F({\bm \theta}) \| p({\bm \theta}|\mathcal{Z})\big)=\E_{q_F({\bm \theta})}\l[\log \l(\frac{q_F({\bm \theta}) }{ p({\bm \theta}|\mathcal{Z})}\r)\r].$$ 
This is the standard objective function used in variational inference (see, e.g., \cite{simeone2017brief, rabinovich2015variational}). This problem is equivalent to the minimization of the variational free energy \cite{josefree}
   \begin{align}\label{eq: max ELBO}
 \min_{\bW}\Big\{ \cL(\bW)& = -\E_{{\bm \theta}\sim q_{F}(\bm \theta)}\big[\log {p({\bm \theta},\mathcal{Z})}\big] -{\sf H}\big[ q_F(\bm \theta)\big] \Big\},
 \end{align}  
which is also known as  negative evidence lower bound (ELBO) \cite{josefree}. Note that the variational free energy  \eqref{eq: max ELBO} does not depend on the unknown posterior $p({\bm \theta}|\cZ)$ but on the joint $p({\bm \theta},\cZ)=p({\bm \theta})p(\cZ|{\bm \theta})=p({\bm \theta})\prod_{k=1}^Kp(\cZ_k|{\bm \theta})$.

In the following, we will develop algorithms that tackle problem \eqref{eq: max ELBO}  for OMA by using the received signals $\{\{\by_k^{(s)}\}_{k=1}^K\}_{s=1}^S$ and for NOMA by using the received signals $\{\by^{(s)}\}_{s=1}^S$. In both cases, the server can also use the data set $\cZ$. The proposed approaches are based on stochastic gradient descent (SGD) applied at the server. Specifically, the server applies SGD steps based on the gradient $\nabla_{\bm \theta} \log p(\bz_n|{\bm \theta})$ for source data point $\bz_n\in \cZ$, which can be readily obtained at the server based on the global data set $\cZ$. In fact, computing gradients $\nabla_{\bm \theta} \log p(\bz_n|{\bm \theta})$ is much less computationally intensive than computing the posterior $p({\bm \theta}|\cZ)$: The latter requires integrating over the space of the model parameters, while the former can be carried out via efficient algorithms such as backpropagation.  For comparison, in Sec \ref{sec: sim}, we will consider the performance of  stochastic gradient Langevin dynamics (SGLD), an MCMC scheme that also requires the computation of the same stochastic gradients.


\subsection{Wireless Variational CMC: OMA}

%
%
%
%
\subsubsection{Approximation of the Objective Function}
For OMA, the global sample ${\bm \theta}\sim q_F({\bm \theta})$ is given as a function of the received signals in \eqref{eq: est theta}. Therefore, the objective in \eqref{eq: max ELBO} can be rewritten as the expectation over the received signals  $\{\by_k\}_{k=1}^K$ as 
   \begin{align} \label{eq: min KL repar}
  \cL(\bW)&= -\E_{\by_1,\cdots,\by_K}\bigg[\log {p\Big( \sum_{k=1}^K \bW_k \by_k,\mathcal{Z}\Big)}\bigg] -{\sf H}\big[ q_F(\bm \theta)\big]. 
 \end{align}  
 This step implements the  ``reparameterization trick" \cite{simeone2017brief}, moving the optimization variables $\bW$ from the averaging distribution to the argument  of the expectation. 
The expectation can now be estimated by averaging over the samples $\{\{\by_k^{(s)}\}_{k=1}^K\}_{s=1}^S$ as
\begin{align}
 \widehat{ \cL}(\bW)& \!= \!-\frac{1}{S}\sum_{s=1}^{S}\Big[\log {p\Big( \sum_{k=1}^K \bW_k \by_k^{(s)},\mathcal{Z}\Big)}\Big]  -  {\sf H}\big[ q_F(\bm \theta)\big].  \!\label{eq: obj sum} 
\end{align}
The  direct minimization of \eqref{eq: obj sum}  is still intractable due to the presence of differential entropy of the distribution $q_F({\bm \theta})$. This issue is addressed via a tractable lower bound, as derived in the following lemma. 
 \begin{lemma}[Lower Bound on Differential Entropy for OMA]\label{lemma: LB entropy}\emph{With given encoding matrices $\{\bE_k\}_{k=1}^K$,  the differential  entropy ${\sf H}[ q_F(\bm \theta)]$ of the distribution $q_F({\bm \theta})$ of the global sample \eqref{eq: est theta} under  OMA is lower bounded~as
 \begin{align}\label{eq: LB entropy}
{\sf H}[ q_F(\bm \theta)] &\geq \frac{d}{2}\log\l(2K\sqrt{2\pi e N_0}\r)+   \frac{1}{2K}\sum_{k=1}^K  \log |\det (\bW_k\bE_k)| \nn\\
& \qquad \qquad  \qquad+{\sf H}[p_k({\bm \theta}_k|\cZ_k) ] +\frac{1}{2}\log \det \l(\bW_k \bW_k^{\sf T }\r)  \nn\\
&:=\widetilde{\sf H}[ q_F(\bm \theta)],
\end{align}
where ${\sf H}[p_k({\bm \theta}_k|\cZ_k) ]$ is the differential entropy of the subposterior  $p_k({\bm \theta}_k|\cZ_k)$.}
\end{lemma}
\proof The proof follows the steps of \cite[Supplement A Proofs]{rabinovich2015variational} and is detailed in Appendix~\ref{proof entropy}.

 Using Lemma~\ref{lemma: LB entropy} in  \eqref{eq: obj sum} yields the following upper bound on the empirical free energy  \eqref{eq: obj sum} for~OMA  
 \begin{align}
\widehat{\cL}(\bW)\leq  \widetilde{\cL}(\bW) := -\frac{1}{S}\sum_{s=1}^{S}\Big[\log {p\Big( \sum_{k=1}^K  {\bW_k \by_k^{(s)}},\mathcal{Z}\Big)}\Big] \nn\\
 -\widetilde{\sf H}\big[ q_F(\bm \theta)\big]. \label{obj OMA}
\end{align}

\subsubsection{Stochastic Gradient Descent Algorithm}
 The proposed strategy, referred to as wireless variational consensus Monte Carlo (WVCMC), aims at minimizing the upper bound \eqref{obj OMA} via SGD. Given a mini-batch $\cB \in \cZ$ of $N_b\leq N$ samples, stochastic gradient of $\widetilde{L}(\bW)$ with respect to weight matrix $\bW_k$ is given  as 
\begin{align}
&\!\!\!\widehat{\nabla}_{\bW_k} \widetilde{\cL}(\bW) \nn\\ 
&\!\!\!=\!-\frac{1}{S}\!\!\sum_{s=1}^{S} \!\!\Big[\!\nabla_{{\bm \theta}^{(s)}}\! \log p({\bm \theta}^{(s)})\!+\! \frac{N}{N_b}\!\!\sum_{\bz_n\in \cB}\!\! \nabla_{{\bm \theta}^{(s)}} \!\!\log p(\bz_n|{\bm \theta}^{(s)}\!)\!  \Big] \!(\by_k^{(s)})\!^{\sf T} \nn\\
& \qquad \qquad \qquad \qquad -\frac{1}{2K} \big[ (\bW_k\bE_k)^{-\sf T}  \bE_k^{\sf  T}+(\bW_k^\dagger)^{\sf T}  \big] ,\label{eq: gradient Dk}
 \end{align}
where we recall that we have the global sample ${\bm \theta}^{(s)}=\sum_{k=1}^K \bW_k \by_k^{(s)}$; and $\bW_k^\dagger $ is the pseudoinverse of $\bW_k$. This yields SGD algorithm based on \eqref{eq: gradient Dk} as summarized in  Algorithm~\ref{algorithm: offline optimization for Dk}.  Note that, unlike GCMC and WGCMC, WVCMC requires the use of data set $\cZ$ at server.  
 \begin{algorithm}
{\bf Input:} Data set $\cZ$, received signals   $\{\by_1^{(s)},\cdots, \by_K^{(s)}\}_{s=1}^{S}$, encoding matrices$\{\bE_k\}_{k=1}^K$, step size $\eta$, number of iterations $t_m$.

{\bf Initialize:}   ${\bW}^0$

{\bf For each iteration:} $t=1,\dots,t_m$ 


\quad  {\bf Estimate samples of global posterior using ${\bW}^{t-1}$:} 

\qquad       ${\bm \theta}^{(s)}=\sum_{k=1}^K{ \bW_k^{t-1}\by_k^{(s)}} $ for $s=1, \dots, S$

\quad  {\bf Update ${\bW}^{t}$:} 

\qquad    Select a mini-batch of data $\cB=\{\bz_n\}_{n=1}^{N_b} \in \cZ$

\qquad     $ {\bW}_k^{t}={\bW}_k^{t-1}-\eta \widehat{\nabla}_{\bW_k} \widetilde{\cL}(\bW^{t-1})  $ 

\qquad   \quad       with $\widehat{\nabla}_{\bW_k} \widetilde{\cL}(\bW^{t-1})$ in \eqref{eq: gradient Dk}
%

{\bf Output:} Global samples ${\bm \theta}^{(s)}=\sum_{k=1}^K{ \bW_k^{t_m}\by_k^{(s)}}$  for $s=1,\dots,S$. 


\caption{WVCMC for OMA}
\label{algorithm: offline optimization for Dk}
\end{algorithm}


{\color{black}In \cite{rabinovich2015variational},  matrices $\{\bW\}_{k=1}^K$ were assumed to be positive semidefinite, resulting in a convex program when the  joint distributions $p({\bm \theta}, \cZ)$  are log-concave in $\bm \theta$. In  the more general setting considered here, the problem is not convex.  The iteration complexity of Algorithm 1 to guarantee convergence to stationary points following directly from standard results on SGD, and it depends on the curvature properties of function $\log p({\bm \theta}, \cZ)$ \cite[Theorem 4.8]{bottou2018optimization}.}


\subsection{Wireless Variational CMC: NOMA}
\subsubsection{Approximation of the Objective Function}
In a manner similar to OMA, we first approximate the objective in \eqref{eq: max ELBO}  by averaging over the received signal $\{\by^{(s)}\}_{s=1}^S$. To this end, we recall that the global samples ${\bm \theta}^{(s)}$ are obtained via \eqref{eq: est theta NOMA}, and hence we can write the empirical free energy \eqref{eq: max ELBO} as 
\begin{align}
\widehat{\cL}(\bW)&\!=\! -\frac{1}{S}\sum_{s=1}^{S}\Big[\log {p\Big(\bW\by^{(s)},\mathcal{Z}\Big)}\Big] -{\sf H}\big[ q_F(\bm \theta)\big].   \label{eq: obj sum noma} 
 \end{align}  
 Again, the direct optimization of \eqref{eq: obj sum noma}  is  intractable given the presence of differential entropy ${\sf H}[q_F(\bm \theta)]$, which is addressed by the tractable lower bound derived below. 
 \begin{lemma}[Lower Bound on Differential Entropy for NOMA]\label{lemma: LB entropy noma}\emph{With given encoding matrices $\{\bE_k\}_{k=1}^K$,  the differential  entropy ${\sf H}[ q_F(\bm \theta)]$ of the distribution $q_F({\bm \theta})$ of the global sample \eqref{eq: est theta NOMA} under  NOMA is lower bounded~as
 \begin{align}\label{eq: LB entropy NOMA}
{\sf H}[ q_F(\bm \theta)] &\geq \frac{d}{2}\log\l[(K+1)\l({2 \pi e N_0}\r)^{\frac{1}{K+1}}\r]  \nn\\
&\quad +\frac{1}{K+1}\bigg(K\log |\det (\bW\bE)|+ \frac{1}{2} \log \det(\bW \bW^{\sf T}) \nn\\
&\qquad \qquad\qquad\qquad \qquad \qquad+ \sum_{k=1}^K   {\sf H}[p_k({\bm \theta}_k|\cZ_k) ] \bigg)   \nn\\ 
&:=\widetilde{\sf H}[ q_F(\bm \theta)].
\end{align}}
\end{lemma}

 Using Lemma~\ref{lemma: LB entropy noma} in  \eqref{eq: obj sum noma} yields the following upper bound on the empirical free energy for NOMA as 
 \begin{align}
\widehat{\cL}(\bW)&\leq  \widetilde{\cL}(\bW) := -\frac{1}{S}\sum_{s=1}^{S}\Big[\log {p\Big(  {\bW \by^{(s)}},\mathcal{Z}\Big)}\Big]  -\widetilde{\sf H}\big[ q_F(\bm \theta)\big]. \label{obj NOMA}
\end{align}

\subsubsection{Stochastic Gradient Descent Algorithm}
The proposed WVCMC algorithm for NOMA aims at   minimizing the upper bound  \eqref{obj NOMA}  via SGD. 
Given a mini-batch $\cB \in \cZ$ of $N_b\leq N$ samples, stochastic gradient of $\widetilde{L}(\bW)$ with respect to weight matrix $\bW$ is given  as 
\begin{align}
&\!\!\widehat{\nabla}_{\bW} \widetilde{\cL}(\bW) \nn\\
&\!\!=\!\!-\frac{1}{S}\!\!\sum_{s=1}^{S}\!\!  \Big[\!\nabla_{{\bm \theta}^{(s)}}\! \log p({\bm \theta}^{(s)})\!+\!\frac{N}{N_b}\!\!\sum_{\bz_n\in \cB}\!\! \nabla_{{\bm \theta}^{(s)}} \!\log \! p(\bz_n|{\bm \theta}^{(s)}) \! \Big]\!  (\by^{(s)})\!^{\sf T}  \nn \\ 
&\qquad \qquad \qquad   -\frac{1}{K+1}\Big[ K (\bW\bE)^{-\sf T}  \bE^{\sf  T}+(\bW^\dagger)^{\sf T}   \Big],\label{eq: gradient Dk noma}
 \end{align}
where we have  ${\bm \theta}^{(s)}= \bW\by^{(s)}$. The resulting algorithm is summarized in  Algorithm~\ref{algorithm: offline optimization for Dk NOMA}. 
 \begin{algorithm}
{\bf Input:} Data set $\cZ$, received signals   $\{\by^{(s)}\}_{s=1}^{S}$, encoding matrices $\bE$, step size $\eta$, number of iterations $t_m$.

{\bf Initialize:}   ${\bW}^0$

{\bf For each iteration:} $t=1,\cdots,t_m$ 

\quad  {\bf Estimate samples of global posterior using  ${\bW}^{t-1}$:} 

\qquad       ${\bm \theta}^{(s)}={ \bW^{t-1}\by^{(s)}} $ for $s=1, \dots, S$

\quad  {\bf Update ${\bW}^{t}$:} 

\qquad    Select a mini-batch of data $\cB=\{\bz_n\}_{n=1}^{N_b} \in \cZ$

\qquad     $ {\bW}^{t}={\bW}^{t-1}-\eta \widehat{\nabla}_{\bW} \widetilde{\cL}(\bW^{t-1})  $   

\qquad \quad with $\widehat{\nabla}_{\bW} \widetilde{\cL}(\bW^{t-1})$ in \eqref{eq: gradient Dk noma}
%

{\bf Output:} Global samples ${\bm \theta}^{(s)}={ \bW^{t_m}\by^{(s)}}$, for $s=1,\dots,S$. 


\caption{WVCMC for NOMA}
\label{algorithm: offline optimization for Dk NOMA}
\end{algorithm}

\section{Experiments}\label{sec: sim}

In this section, we evaluate the performance of the proposed channel-driven CMC schemes for two standard problems: 1) CMC with Gaussian subposteriors \cite{scott2016bayes}; and 2) CMC for Bayesian probit regression on  synthetic data set and  MNIST data set. To this end, we first introduce the performance metrics and reference schemes that are adopted throughout this section.

\subsubsection{Performance Metrics}
As in prior works on CMC \cite{rabinovich2015variational}, the performance of a CMC algorithm is evaluated by measuring the relative difference between two sample averages of a test function $f({\bm \theta})$, namely the reference average obtained under the true global posterior \eqref{eq: global posterior} and the average computed using samples produced by the CMC algorithm.  Given a test function $f(\cdot)$, denoting ${\bm \theta}_g^{(s)}\sim p({\bm \theta}|\cZ)$ for $s=1,\dots,S_g$ as a set of samples from  the distribution of global posterior \eqref{eq: global posterior},  the test error is defined as 
\begin{align}
{\sf err}(f(\cdot))=\frac{\big|\frac{1}{S}\sum_{s=1}^{S} f( {\bm \theta}^{(s)}) -  \frac{1}{S_g}\sum_{s=1}^{S_g} f( {\bm \theta}_g^{(s)}) \big|}{\big| \frac{1}{S_g}\sum_{s=1}^{S_g} f( {\bm \theta}_g^{(s)}) \big|},  \label{eq: eval func}
\end{align}
where $\{{\bm \theta}^{(s)}\}_{s=1}^S$ are the samples produced by the CMC algorithm under evaluation. 
Specifically, as in  \cite{rabinovich2015variational}, we consider multiple test functions, with each function given by one entry of the outer product matrix ${\bm \theta}{\bm \theta}^{\sf T}$. Accordingly, we have the test functions $f_{ij}(\bm \theta)= {\bm \theta}[i]  {\bm \theta}[j]$ for $i,j=1,\dots, d$, where $ {\bm \theta}[i]$ is the $i$-th entry in vector ${\bm \theta}$. We then average the per-test function error \eqref{eq: eval func} to obtain 
\begin{align}\label{eq: err2}
{\sf err}=\frac{1}{d^2}\sum_{i=1}^d \sum_{j=1}^d  \frac{\big|\frac{1}{S}\sum_{s=1}^{S} f_{ij}({\bm \theta}^{^{(s)}}) -  \frac{1}{S_g}\sum_{s=1}^{S_g}f_{ij}({\bm \theta}_g^{^{(s)}})  \big|}{\big|\frac{1}{S_g}\sum_{s=1}^{S_g} f_{ij}({\bm \theta}_g^{^{(s)}})\big|}. 
\end{align}
Accordingly,  the error \eqref{eq: err2} quantifies the accuracy of the empirical correlation matrix of the posterior distribution that can be obtained from the samples $\{{\bm \theta}^{(s)}\}_{s=1}^S$. Similar results are obtained for the estimate of other moments of the posterior such as the mean vector. 

{\color{black}
We will also consider the inference performance of the ensemble predictive distribution 
\begin{align} 
\widehat{p}_{\{{\bm \theta}^{(s)}\}_{s=1}^S}(v  |\bu, \cZ)=\frac{1}{S}\sum_{s=1}^S p(v|{\bm \theta}^{(s)},\bu,\cZ) \label{eq: pred dist1} 
\end{align} 
obtained at the server given the samples $\{{\bm \theta}^{(s)}\}_{s=1}^S$ for any test input $\bu$. Specifically, the quality of predictive distribution \eqref{eq: pred dist1}  is measured by the Kullback-Leibler (KL) divergence between \eqref{eq: pred dist1}  and the ideal ensemble distribution $\widehat{p}_{\{{\bm \theta}_g^{(s)}\}_{s=1}^S}(v  |\bu, \cZ) $ computed by using the samples of true posterior distribution, i.e., 
\begin{align}\label{eq: KL divergence}
\frac{1}{N_t}\sum_{i=1}^{N_t}{\rm KL}\Big(\widehat{p}_{\{{\bm \theta}^{(s)}\}_{s=1}^S}(v  |\bu_i, \cZ) \Big \| \widehat{p}_{\{{\bm \theta}_g^{(s)}\}_{s=1}^S}(v  |\bu_i, \cZ) \Big), 
\end{align} 
where $\{\bu_i\}_{i=1}^{N_t}$ is a given test set. 
}


\subsubsection{Reference Schemes}
We evaluate the performance of:  (\emph i) GCMC in  \eqref{eq: est GCMC result} with decoded received signals $\{\bE^\dagger  \by_k^{(s)} \}_{s=1}^S$  in lieu of $\{{\bm \theta}_k^{(s)}\}_{s=1}^S$; (\emph{ii}) WGCMC in  \eqref {eq: est WGCMC result} for OMA and in \eqref{eq: est WGCMC result NOMA} for NOMA; and (\emph{iii}) WVCMC in Algorithm 1 and Algorithm~2.  Furthermore, in order to evaluate the advantages of leveraging the $K$ workers in the wireless data center, we also implement a state-of-the-art scheme for MC sampling at the server, namely SGLD \cite{welling2011bayesian}.  SGLD is selected because it can be run with the same order of computational complexity at the server of the proposed WVCMC schemes. In fact, as detailed next, SGLD relies on the computation of stochastic gradients $\nabla \log p({\bm \theta}, \bz_n)$ for randomly selected examples $\bz_n\in \cZ$. Therefore, we can compare SGLD and WVCMC under the assumption of (approximately) equal computation load at the server. 

At each iteration $t$, SGLD updates the model parameter vector by using a subset of data in the global data set $\cZ$ as 
 \begin{align} 
\text{(SGLD)} \quad  {\bm \theta}^{(t+1)}\!=\!{\bm \theta}^{(t)}\!\!+\!\!\frac{\eta^{(t)}}{2} &\Big[ \!\frac{N}{N_b}\sum_{\bz_n\in \cB} \nabla_{{\bm \theta}^{(t)}} \log p(\bz_n|{\bm \theta}^{(t)})  \nn\\
&\!\!+\!\nabla_{{\bm \theta}^{(t)}}\! \log p({\bm \theta}^{(t)})\!\Big]\!\!+\!\bn^{(t)}\!,\! \label{eq: update SGLD}
 \end{align}
 where the learning rate is set as $\eta^{(t)} =\alpha(\beta+t)^{-\gamma}$ for some $\beta>0$ and $\gamma \in (0.5,1]$, and $\bn^{(t)} \sim\cN(0, \eta^{(t)} \bI)$ is injected noise. SGLD is known to produce, in the limit of a large number of iterations, samples from the desired posterior. After removing the first $t_b$ samples, which we consider to constitute the burn-in phase, the remaining $\{{\bm \theta}^{(t)}\}_{t=t_b+1}^{t_m}$ samples are used for evaluation in \eqref{eq: err2}.  We emphasize that SGLD does not make use of the presence of the $K$ workers. Distributed versions of SGLD also exist. These can be useful as alternatives to the CMC schemes studied in this paper when the workers do not have sufficient computing power to obtain samples from the local subposteriors -- a condition that we have assumed. We refer to Sec. \ref{sec: conclusions} for more discussion on this point. As for the  computation at the server, consider $C$ computed gradients $\nabla_{{\bm \theta}} \log p({\bm \theta},\bz_n)$, SGLD requires $C/N_b$ iterations while  WVCMC requires $C/N_bS$ iterations since WVCMC process $S$ samples $\{{\bm \theta}^{(s)}\}_{s=1}^S$ in parallel.

\subsection{Gaussian Subposteriors} 

We start with the baseline example studied in Sec. \ref{sec: GCMC} and Sec. \ref{sec: WGCMC}, in which the local subposterior for each worker $k$ follows a Gaussian distribution $\cN(0,\bC_k)$. The model dimension is $d=5$, and each covariance matrix $\bC_k$ is a symmetric Toeplitz matrix with first column given as  $[ 1, \rho_k,    \rho_k^2,   \rho_k^3,  \rho_k^4]$ for some $\rho_k \in [-1,1] $. We consider $K = 10$ workers, and set $\rho_k=(k-1)/K$ for $k$-th worker.  For reference, we also study a homogeneous setting in which the covariances are equal to  $\bC_0=K( \sum_{k=1}^K \bC_k^{-1})^{-1}$ for all workers. 

To start, we consider frequency-flat channels with $\bH_k=\bI_d$, and the $\SNR=P/(dN_0)$ is set as $5$ dB by default. Unless stated otherwise, the number of communication blocks is equal to $T=2000$; the number of per-output-sample iterations  is $t_m=30$ for NOMA and $t_m=300$ for OMA, which ensures that the same total number of iterations, namely $300$, are carried out for both OMA and NOMA;   the learning rate is $\eta=10^{-3}$ for NOMA and $\eta=5 \times 10^{-3}$ for OMA, which have been separately optimized through a grid search;  and the initial point used for WVCMC is $\bW^0=1/K\bI_d$ for NOMA, while in OMA it is set as GCMC solution. This initialization given by the GCMC solutions was found to be advantageous for OMA, but it is not applicable to NOMA.  Note that, in this simple example, the global posterior covariance matrix is  $\bC=( \sum_{k=1}^K \bC_k^{-1})^{-1}$, and hence we use this ground-truth matrix in \eqref{eq: err2} in lieu of the estimate $\frac{1}{S_g}\sum_{s=1}^{S_g}{\bm \theta}_g^{{(s)}}( {\bm \theta}_g^{{(s)}})^{\sf T} $. We use full mini-batches, i.e., we implement gradient descent, such that the gradients $\nabla_{\bm \theta} p({\bm \theta},\cZ)$ in  \eqref{eq: gradient Dk}  and \eqref{eq: gradient Dk noma} are  computed as $\nabla_{\bm \theta} p({\bm \theta},\cZ)=\nabla_{\bm \theta} p({\bm{\theta}}|\cZ) $. 
All results are averaged over $100$ experiments.  

\subsubsection{Error vs number of communication blocks} 

\begin{figure}[t]
\centering
\includegraphics[width=7.5cm]{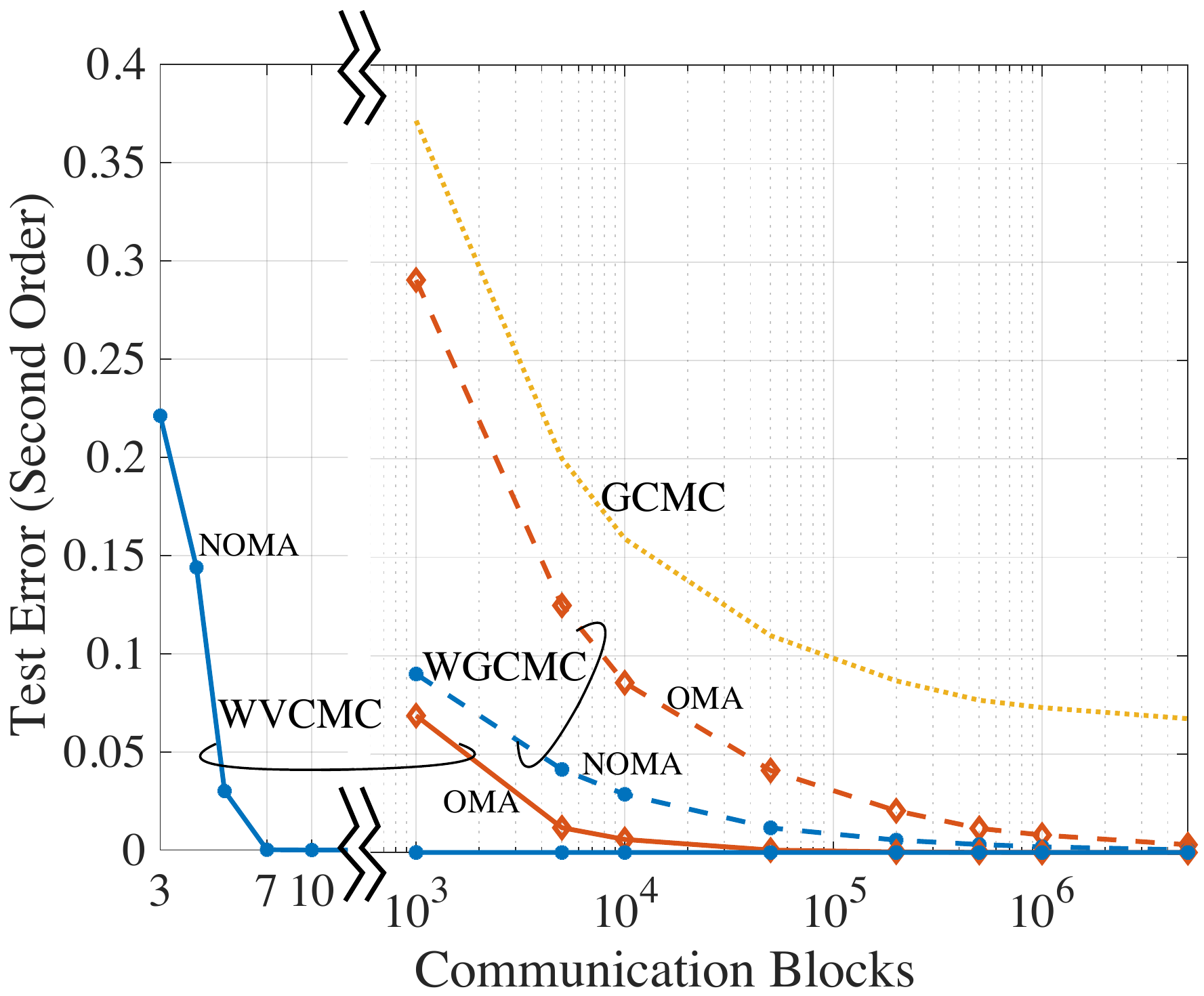}
\caption{Second-order test error \eqref{eq: err2} versus the number of communication block, $T$, for wireless CMC ($\SNR=5$~dB, homogeneous subposteriors).}
\label{Fig: comm block}
\end{figure}

In Fig. \ref{Fig: comm block}, we consider homogeneous subposteriors, and plot the test error \eqref{eq: err2} versus the number of communication blocks. We recall that the number of samples is $S=T/K$ for OMA and  $S=T$ for NOMA.  
WVCMC with NOMA is seen to be the most communication-efficient approach, as its test error vanishes for a number of blocks $T\simeq 7$,  while the other schemes require a large number of communication blocks. Validating the analysis in Sec. \ref{sec: WGCMC},  WGCMC, under both OMA and NOMA, is observed to attain  optimality as the number of communication blocks, and hence of samples, increases. We also note that, as mentioned in Sec. \ref{sec: WGCMC}, the conventional GCMC scheme is not asymptotically optimal with a large number of communication blocks. 

\subsubsection{Error vs SNR} 
\begin{figure}[t]
\centering
\includegraphics[width=7.5cm]{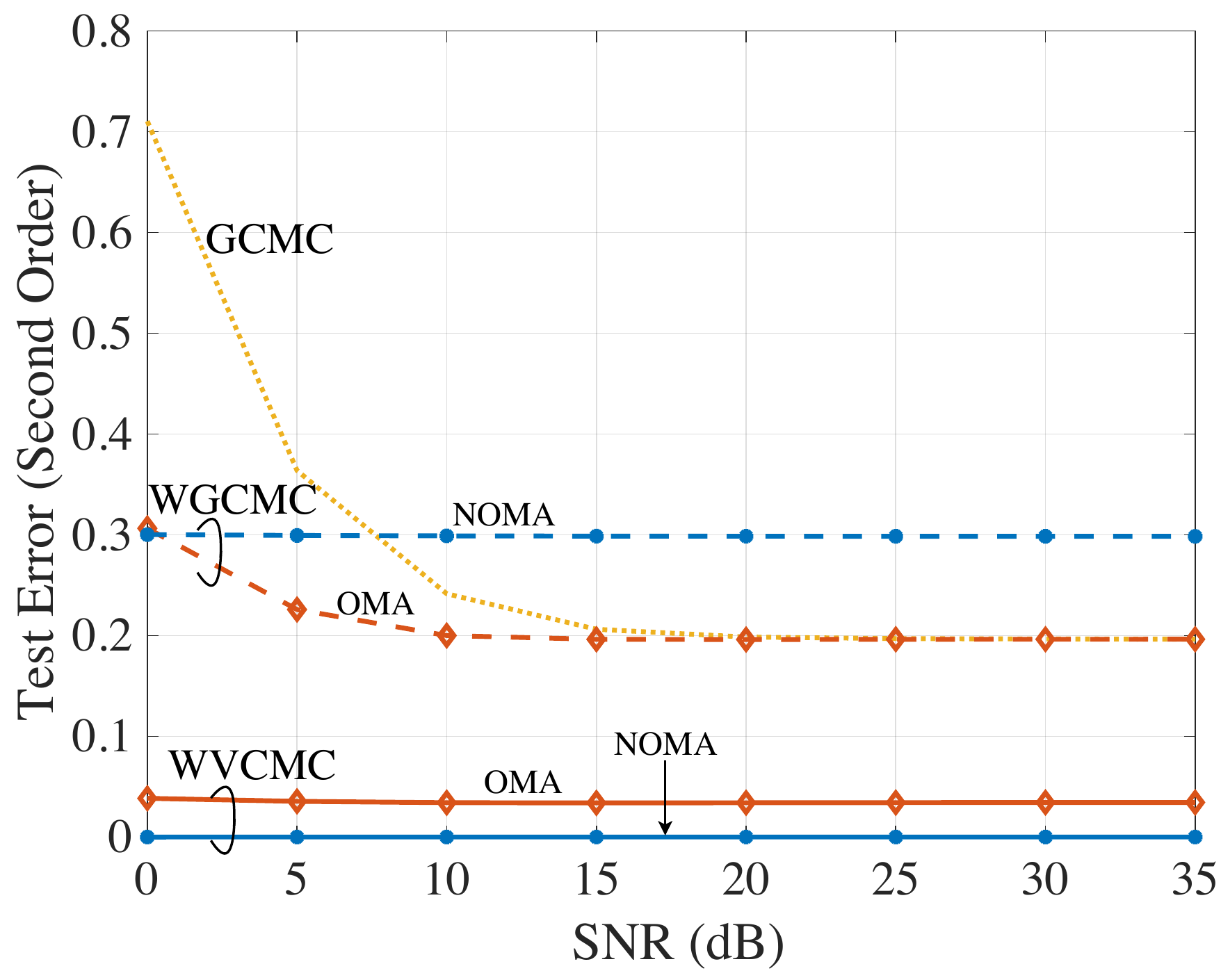}
\caption{Second-order test error \eqref{eq: err2} versus SNR for wireless CMC ($T=2000$, heterogeneous subposteriors).}
\label{Fig: snr}
\end{figure}

We now study the impact of SNR in Fig. \ref{Fig: snr} under the assumption of heterogeneous subposteriors, i.e., subposteriors with different  covariance matrices $\{\bC_k\}_{k=1}^K$ as described above.  WVCMC is seen to successfully adapt to different SNR levels, outperforming all the other schemes. In fact, the error of WVCMC is essentially unaffected by the SNR level in the considered range. This observation points to a key property of the proposed channel-driven MC sampling: If properly accounted for, the randomness injected by the channel noise may not affect the performance of CMC algorithms. Channel noise can be considered as a component of the sample generation process, rather than a nuisance to be mitigated. 
Importantly, the results in Fig. \ref{Fig: snr} suggest that it is critical to properly optimize the aggregation function in order to leverage this unique property of MC sampling over noisy channels. We also note that the noise-invariant error of WVCMC is attained in the ideal case studied in this example, for which the subposterior and channel noise follow Gaussian distributions. 
As two further comments on Fig. \ref{Fig: snr}, WGCMC under OMA is confirmed to be  advantageous over conventional GCMC, with the performance of two schemes converging at high SNR. Moreover, WGCMC under NOMA is seen to be worse than that of OMA, due to the performance loss counted by the heterogeneity of the subposteriors (see Sec. \ref{sec: WGCMC noma} for further discussion). Note that this is unlike the proposed WVCMC NOMA scheme, which remains advantageous over OMA.

\subsection{Bayesian Probit Regression}\label{sec: sim synthetic data}
We turn to focus on the more complex problem of Bayesian probit regression. Accordingly, given observations $\{\bz_n=(\bu_n,v_n)\}_{n=1}^N$, the likelihood is given by the following binary regression model
\begin{align}\label{eq: probit model}
p(v_n=1|{\bm \theta},\bu_n)=\Phi({\bm \theta}^{\sf T} \bu_n),
\end{align}
where  $\Phi(\cdot)$ is the standard normal cumulative distribution function. 
Furthermore, the prior is the Gaussian distribution $\cN(0,\sigma^2 \bI)$ for a fixed variance $\sigma^2$.

We first consider a synthetic data set following the model described in \eqref{eq: probit model}. This is comprised of $N=8500$ labeled data points $\{\bz_n=(\bu_n,v_n)\}_{n=1}^N$. The covariates $\bu_n \in \mathbb{R}^{d}$ are drawn i.i.d. from Gaussian distribution $\mathcal{N}(0, \bI_d)$ where $d=5$, and the binary label $v_n$ is drawn as per \eqref{eq: probit model} under the ground-truth model parameter ${\bm \theta}^*=[0.1103, -0.5832, 0.6417, 1.8279, 0.4968]^{\sf T}$. {\color{black}Following \cite{rabinovich2015variational},  we also consider a higher-dimensional data set comprised by $N=15000$ data points with dimension $d=300$ generated in the same manner.}

The channels $\bH_k$ are  $2d \times (2d+2)$ random matrices with i.i.d  standard Gaussian entries, and thus we have $\E\big[ \big(\bH_k \bH_k^{\sf T})\big]^{-1}=\bI_{2d}$.  We implement analog repetition coding, i.e., we set the encoding matrices as $\bE_k=\sqrt{P_k}[\bI_d, \bI_d]^{\sf T}$ for OMA and $\bE=\min_k \sqrt{P_k}[\bI_d, \bI_d]^{\sf T}$ for NOMA, where  $P_k= PS/\sum_{s=1}^S2 \|{\bm \theta}_k^{(s)}\|^2$.  Unless stated otherwise, the data set is divided equally among $K=20$ workers; 
the number of communication blocks is set to $T=1000$; 
the number of per-output sample iterations  to $t_m=50$ for NOMA and OMA;
 the learning rate to $\eta=10^{-7}$ for NOMA and $\eta=10^{-6}$ for OMA upon separate optimizations;
 and the initial point used for WVCMC is $\bW^0=1/K\E_k ^\dagger$ for NOMA while in OMA it is set as GCMC solution as in the previous example. 
 We use full mini-batches, i.e., $N_b=N$ in  \eqref{eq: gradient Dk}  and \eqref{eq: gradient Dk noma}, and details about gradient computation and about the implementation of Gibbs sampling from (sub)posteriors are given in Appendix \ref{appendix: implementation}. To show the effect of the partitions of local data set and the necessity of consensus, we also include the performance of the best single worker in OMA for reference. In this case, the server only uses the samples from the worker that yields the smallest test error.  
Unless stated otherwise, the results  are averaged over $40$ experiments.

\subsubsection{Computational Efficiency at the Server} 
\begin{figure}[t]
\centering
\includegraphics[width=7.5cm]{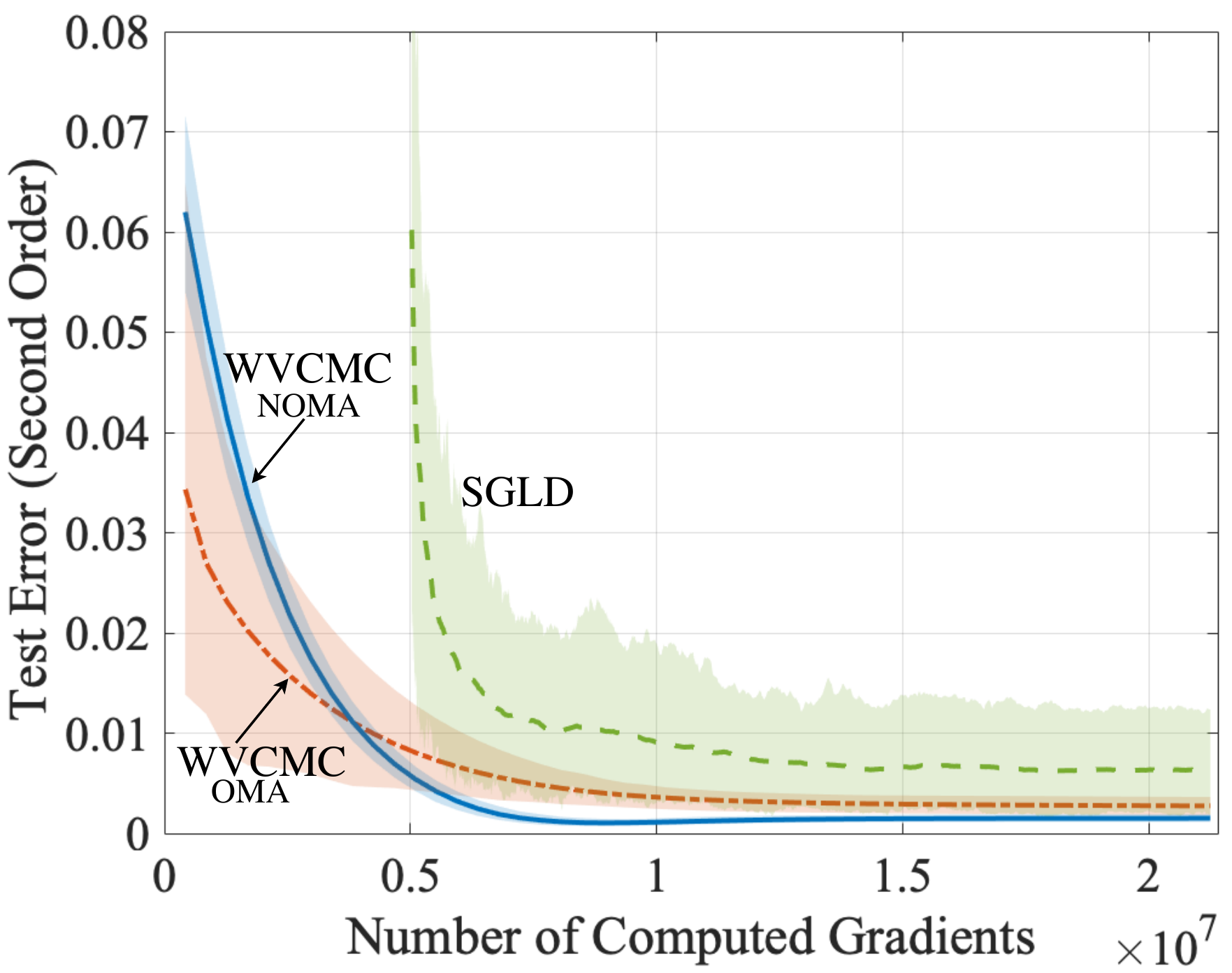}
\caption{Second-order test error \eqref{eq: err2} versus the number of computed gradients $\nabla_{\bm \theta}\log p({\bm \theta},\bz_n)$ for wireless VCMC ($\SNR=15$~dB, $S=50$) and centralized SGLD.}
\label{Fig: CompEfficiency}
\end{figure}

In this experiment, we compare different solutions as a function of the total number of  computed gradients at the server. As discussed, we are specifically interested in the comparison with SGLD. In Fig. \ref{Fig: CompEfficiency}, we plot the test error  \eqref{eq: err2} versus the number of computed gradients $\nabla_{\bm \theta}\log p({\bm \theta},\bz_n)$, with the shaded area representing $90\%$ high-probability interval to illustrate both mean and variance during the whole training process. We recall that the number of computed gradients at the server is $NSt$ for WVCMC and $N_b t$ for SGLD where $t$ is the iteration round. Note that we highlight high-probability intervals here and not in other figures since the other plots only show the performance at convergence. 

For WVCMC, we use $S=50$ samples, corresponding to the number of communication blocks  $T=1000$ for OMA and $T=50$ for NOMA, and the $\SNR=P/(2d N_0)$ is set as $15$ dB. For SGLD,  the initialization is randomized using a sample from the prior;  mini-batch size in \eqref{eq: update SGLD} is $N_b=500$; the learning rate is  $\eta^{(t)}= \alpha(\beta+t)^{-\gamma}$, with $\alpha=0.01$, $\beta=1$, $\gamma=0.7$; and number of burn-in samples $t_b=10^4$. All results are averaged over $100$ experiments. 

The most important observation is that, by leveraging the computing power of $K$ parallel workers, WVCMC can vastly outperform SGLD for the same number of computed gradients at the server. 
WVCMC with NOMA is seen to achieve the best accuracy when the number of computed gradients are larger than $5\times10^6$, and such a performance is attainable with least variance among all the schemes. The performance of WVCMC with OMA has a larger variance which is caused by the additive channel noise accumulated across multiple communication blocks due to the more inefficient orthogonal access. 

\subsubsection{Error vs SNR} 
\begin{figure}[t]
\centering
\includegraphics[width=7.5cm]{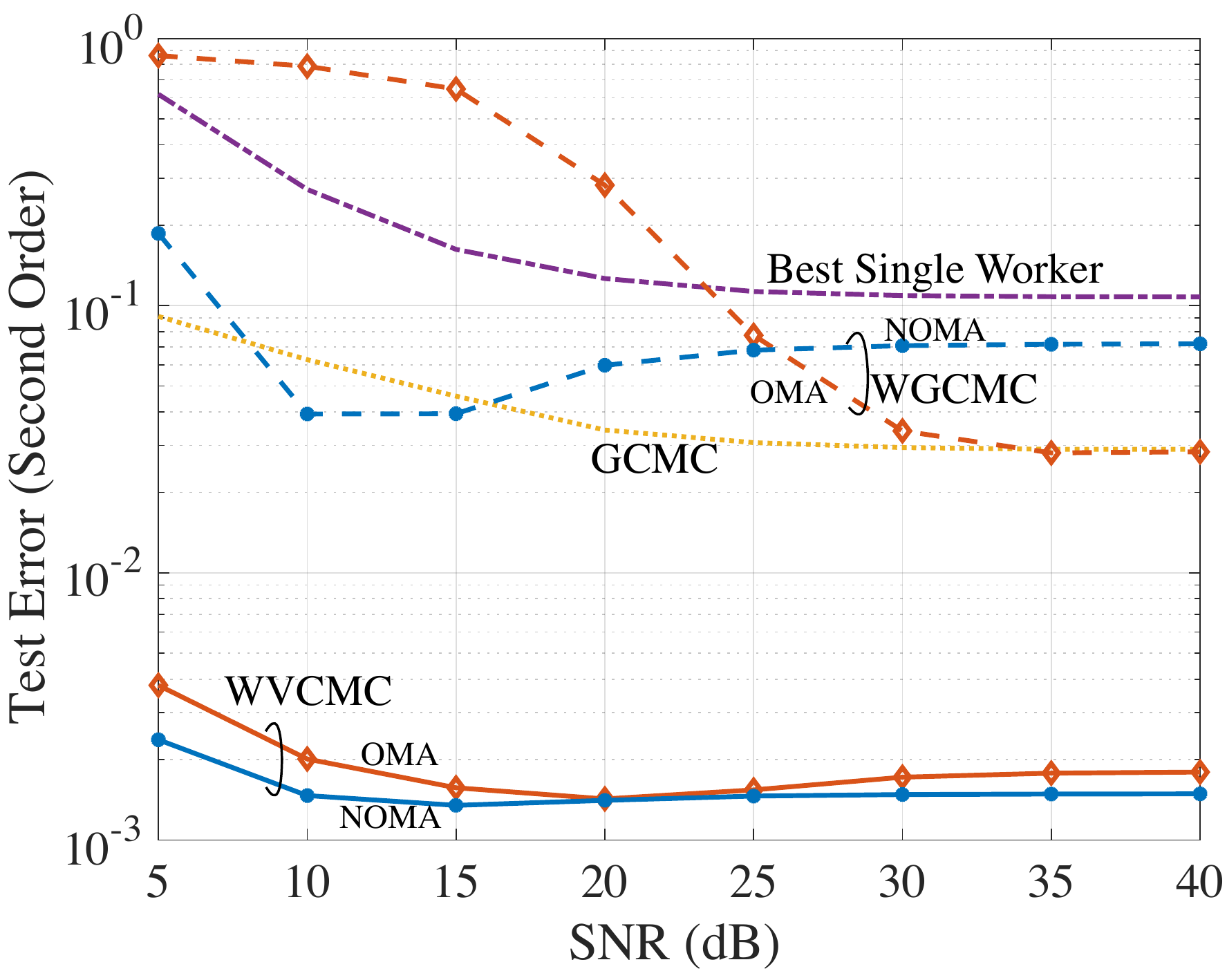}
\caption{Second-order test error \eqref{eq: err2} versus SNR for wireless CMC schemes ($T=1000$, $K=20$).}
\label{Fig: snr probit}
\end{figure}

Fig. \ref{Fig: snr probit} plots the test error in \eqref{eq: err2} versus the SNR. 
WVCMC is again confirmed to outperform all the other schemes. However, unlike the Gaussian example, for which the performance of WVCMC is unaffected by the SNR level, for the non-Gaussian subposteriors considered here, the performance degrades slightly in the low-SNR regime. Unlike WVCMC, the performance of WGCMC schemes is significantly decreased by the non-Gaussianity of the subposteriors. 

\subsubsection{Error vs Number of workers} 
\begin{figure}[t]
\centering
\includegraphics[width=7.5cm]{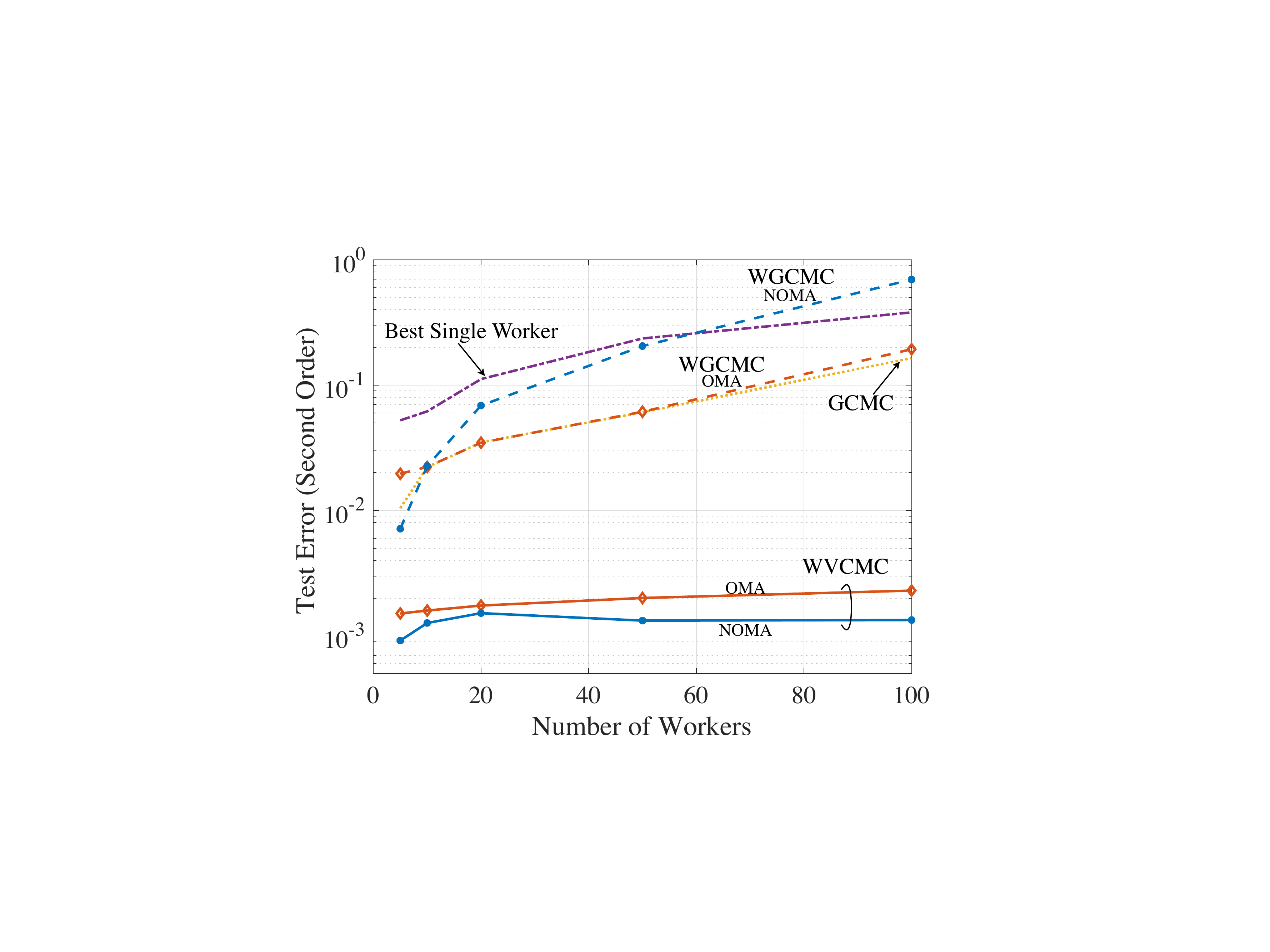}
\caption{Second-order test error \eqref{eq: err2} versus the number of workers for wireless CMC ($T=1000$, $\SNR=35$ dB).}\label{Fig: NumWorker}
\end{figure}

We now study the impact of the number of workers $K$ for $\SNR=35$ dB. The learning rate is set to adapt to the number of workers as $\eta=2\times10^{-5}/K$ for OMA and  $\eta=2\times10^{-6}/K$ for NOMA.  
Fig. \ref{Fig: NumWorker} shows that increasing the number of workers has a negative effect on all the schemes. This degradation is, in part, caused by the reduction in the size of local data sets, and by the resulting need to combine more local samples \cite{rabinovich2015variational,scott2016bayes}. For OMA, increasing $K$ has the addition drawback of reducing the number of received samples for, which is seen to further degrade its performance for $K\geq20$.

\subsubsection{Data Heterogeneity}
\begin{figure}[t]
\centering
\includegraphics[width=7.5cm]{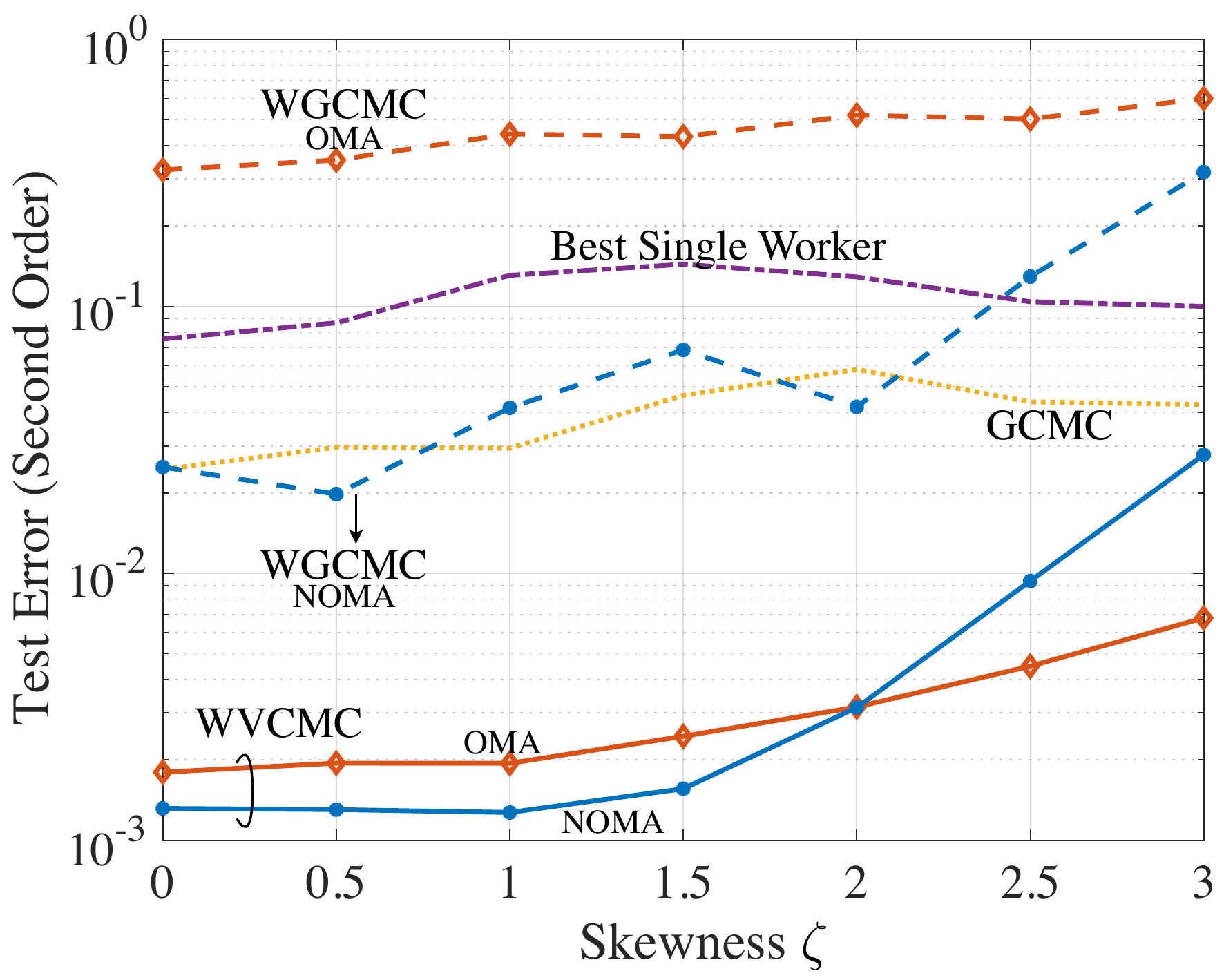}
\caption{Second-order test error \eqref{eq: err2} versus the skewness of data heterogeneity for wireless CMC ($T=1000$, $\SNR=20$ dB).}
\label{Fig: Het}
\end{figure}

Fig. \ref{Fig: Het} plots the error versus a measure of the heterogeneity of the local data sets. To this end, the global data set is partitioned as follows. The $k$-th worker contains a fraction $\frac{1/k^{\zeta }}{\sum_{k=1}^K 1/k^{ \zeta }  } $ of data points from class `1' and a fraction $\frac{1/(K-k+1)^{\zeta}}{\sum_{k=1}^K 1/k^{ \zeta}  } $ of data points from class `0'.  The heterogeneity is described by the exponent  $\zeta$, with $\zeta=0$ representing homogeneous local data sets. As shown in Fig. \ref{Fig: Het}, 
increasing the heterogeneity of local data sets generally affects negatively to all schemes. Nevertheless, the heterogeneity of local data sets has a stronger impact for NOMA than for OMA, since, as discussed in Sec. \ref{sec: WGCMC noma}, NOMA ideally requires  homogeneous subposteriors to achieve optimality. 

{\color{black}
\subsubsection{Higher-Dimensional Data}
\begin{figure}[t]
\centering
\includegraphics[width=7.5cm]{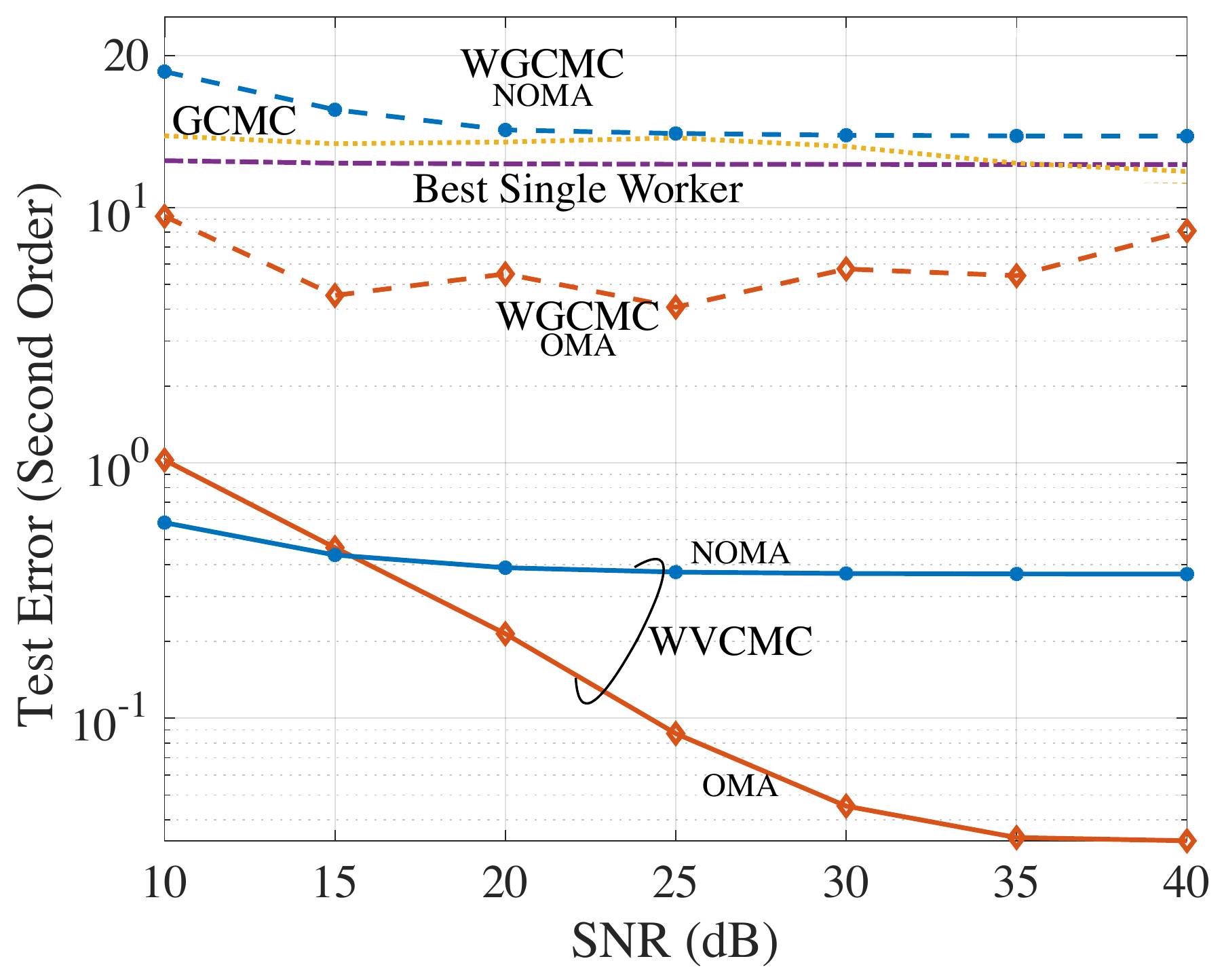}
\caption{Second-order test error \eqref{eq: err2} versus SNR for wireless CMC schemes on high-dimensional data ($T=10000$, $K=20$).}
\label{Fig: snr high dim}
\end{figure}

 We now consider the problem of Bayesian probit regression on higher-dimensional data, matching the scale of the experiments in \cite{rabinovich2015variational}.  The number of  iterations is selected as a function of the SNR  level  to guarantee the convergence, and the learning rate is set to $\eta=5\times10^{-9}$ for NOMA and $\eta=10^{-7}$ for OMA.  All the other settings are same as in the rest of this subsection, and the results are averaged over three realizations. Fig. \ref{Fig: snr high dim} plots the test error in \eqref{eq: err2} versus the SNR.  Confirming the results in Fig. \ref{Fig: snr probit}, WVCMC is seen to significantly outperform all the other schemes. Furthermore, as compared to the case of lower-dimensional data, OMA is seen to be generally advantageous with respect to NOMA, suggesting that the restriction to equal weights matrices in NOMA becomes more problematic as the problem dimension increases.

}

\subsection{MNIST Data Set}
We now consider the problem of Bayesian probit regression on the MNIST data set by using  $N=12665$  data points from the classes of $0$ and $1$. The original  data with dimension $d=784$ is projected to the lower dimension $d=30$ via principal component analysis (PCA). {\color{black}We note that the application of Bayesian learning to lower-dimensional data embeddings is a common approach to enhance scalability \cite{binois2020choice, nayebi2019framework, moriconi2020high, kristiadi2020being}.}
The prior distribution, channel statistics, and initial points used for learning are same as in Sec. \ref{sec: sim synthetic data}.  We use the mini-batch size $N_b=N/20$ in  \eqref{eq: gradient Dk}  and \eqref{eq: gradient Dk noma}. 
Unless stated otherwise, the data set is divided equally among $K=10$ workers; 
the number of communication blocks is set to $T=500$; 
the number of per-output sample iterations  to $t_m=250$ for NOMA and OMA;
and  the learning rate to $\eta=2\times 10^{-7}$ for NOMA and $\eta=10^{-6}$ for OMA upon separate optimizations.

\subsubsection{Computational Efficiency at the Server} 
\begin{figure}[t]
\centering
\includegraphics[width=7.5cm]{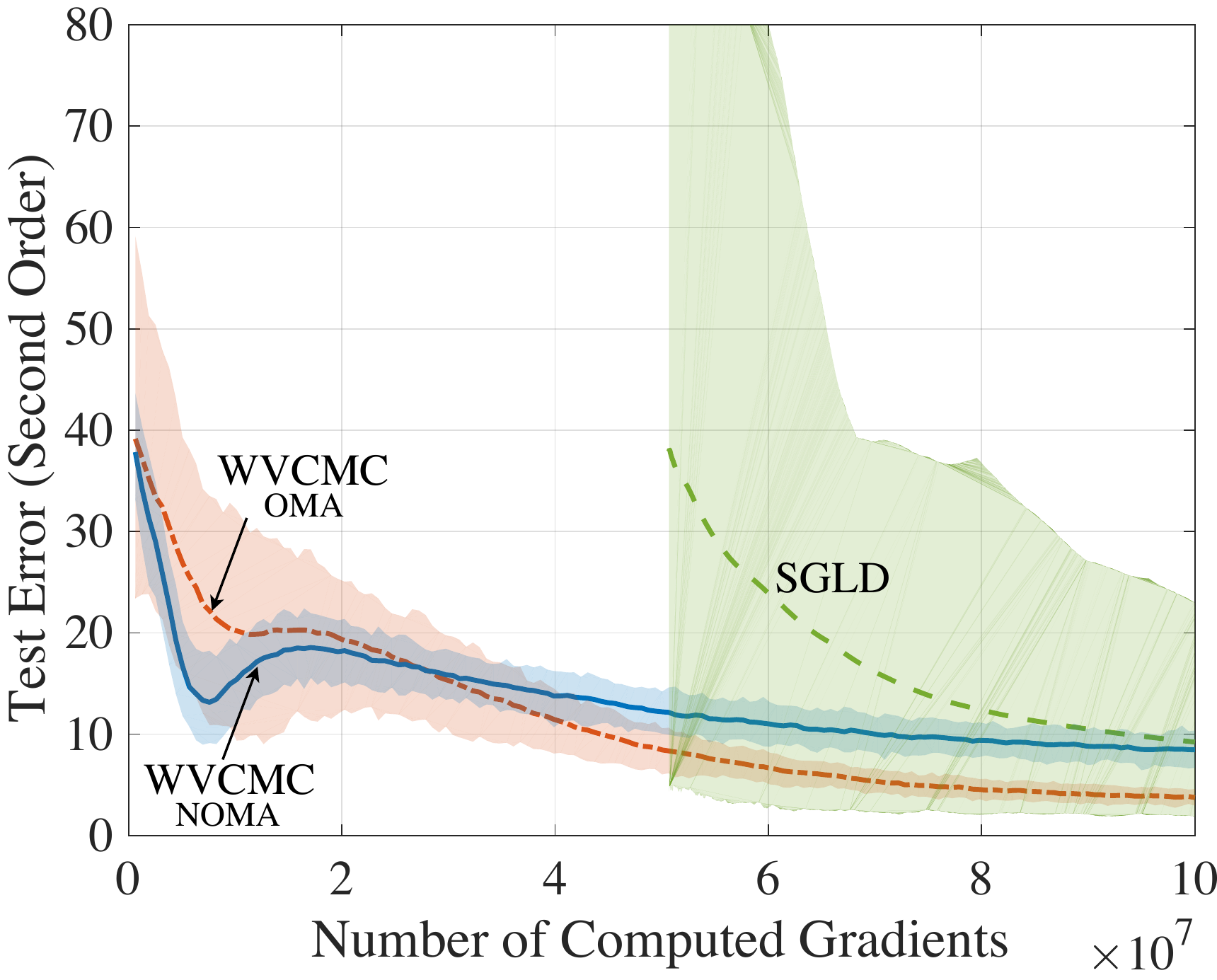}
\caption{Second-order test error \eqref{eq: err2} versus the number of computed gradients $\nabla_{\bm \theta}\log p({\bm \theta},\bz_n)$ for wireless VCMC ($\SNR=30$~dB, $S=50$) and centralized SGLD on MNIST dataset.}
\label{Fig: CompEfficiency probit MNIST}
\end{figure}

We start by validating the experiment in Fig.  \ref{Fig: CompEfficiency} on this data set. To this end, in this experiment, we use $S=50$ samples for WVCMC, corresponding to the number of communication blocks  $T=500$ for OMA and $T=50$ for NOMA, and the $\SNR=30$ dB.  For SGLD, mini-batch size in \eqref{eq: update SGLD} is $N_b=N/20$; the learning rate is  $\eta^{(t)}= \alpha(\beta+t)^{-\gamma}$, with $\alpha=0.01$, $\beta=1$, $\gamma=0.52$; and number of burn-in samples $t_b=8\times10^4$. All results are averaged over $100$ experiments. In Fig. \ref{Fig: CompEfficiency probit MNIST}, we plot the test error  \eqref{eq: err2} versus the number of computed gradients $\nabla_{\bm \theta}\log p({\bm \theta},\bz_n)$, with the shaded area representing $90\%$ high-probability interval. Confirming the results in Fig. \ref{Fig: CompEfficiency}, WVCMC can vastly outperform SGLD for the same number of computed gradients at the server. However, unlike in the experiment on synthetic data set, for MNIST,  WVCMC with OMA is seen to achieve the best accuracy in terms of mean and variance performance. This is due to the heterogeneity of the subposteriors entailed by the use of real data, which degrades the performance of NOMA.    

\subsubsection{Error vs SNR} 

\begin{figure}[t]
\centering
\includegraphics[width=7.5cm]{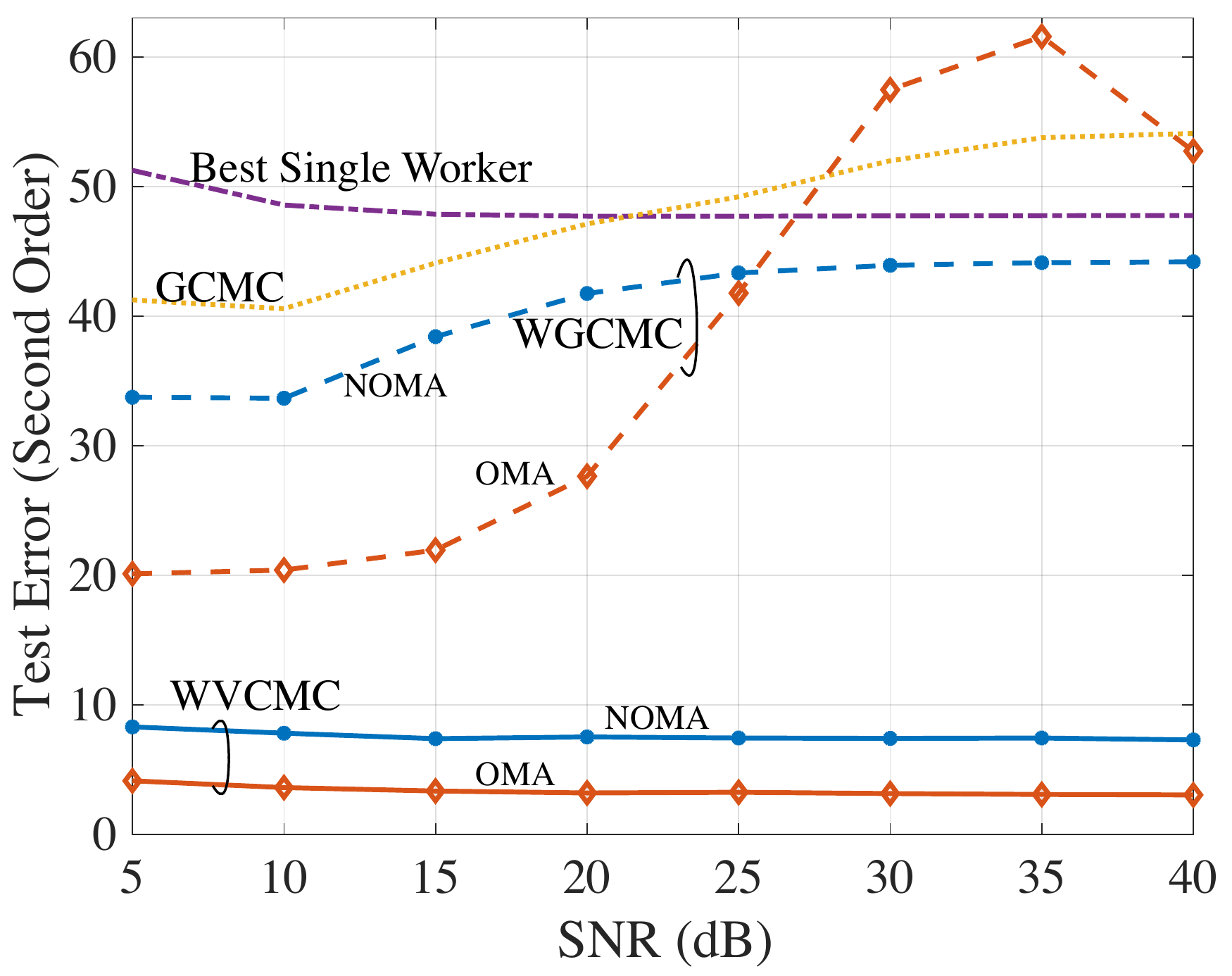}
\caption{Second-order test error \eqref{eq: err2} versus SNR for wireless CMC schemes on MNIST dataset ($T=500$, $K=10$). }
\label{Fig: snr MNIST}
\end{figure}

Finally, Fig. \ref{Fig: snr MNIST} plots the test error in \eqref{eq: err2} versus the SNR.  
WVCMC is confirmed to outperform all the other schemes, and OMA outperforms NOMA due to the heterogeneous subposteriors. Note that, for GCMC and WGCMC, the error is seen to increase in high SNR regime, suggesting that the presence of Gaussian noise in low SNR is actually advantageous to schemes that are designed under the assumption of Gaussian subposteriors.  Furthermore, the limited received samples in OMA cause the inaccurate estimation of covariance matrix, and WGCMC with OMA and GCMC is seen to perform worse than the single-worker benchmark.  

{\color{black}
\subsubsection{Evaluation of Inference Performance} 

\begin{figure}[t]
\centering
\includegraphics[width=7.5cm]{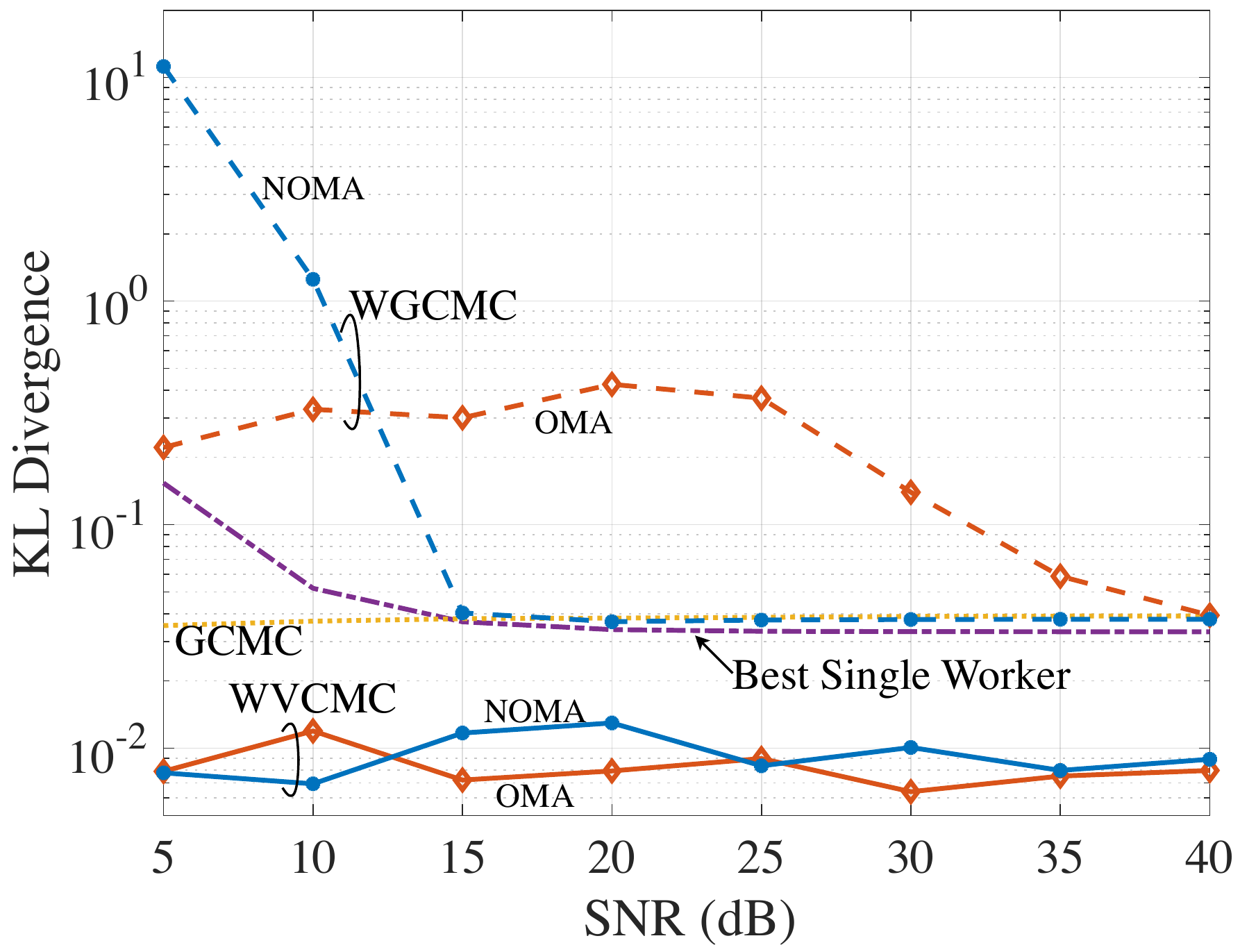}
\caption{KL divergence \eqref{eq: KL divergence} measuring the accuracy of the ensemble prediction \eqref{eq: pred dist1} versus SNR for wireless CMC schemes on MNIST dataset ($T=500$, $K=10$). }
\label{Fig: snr KL}
\end{figure}


We now study the inference performance of the ensemble prediction \eqref{eq: pred dist1} via the KL divergence \eqref{eq: KL divergence} evaluated on a test set of $N_t=2115$ PAC-pre-processed MNIST images. The results in Fig. \ref{Fig: snr KL} confirm the conclusions shown from Fig. \ref{Fig: snr probit} and Fig. \ref{Fig: snr high dim} in terms of the advantages with respect to all benchmarks. 


}

\section{Conclusions}\label{sec: conclusions}

In this paper,  we have considered, for the first time, consensus Monte Carlo (CMC) solutions for distributed one-shot  Bayesian learning in wireless data centers.  Two main ideas were leveraged by the proposed designs. The first is the novel concept of channel-driven sampling: under uncoded wireless transmission of  samples from workers to server, channel noise can contribute to MC sampling, and, if properly accounted for,  is not necessarily a nuisance to be mitigated. In particular, we have argued that channel-driven sampling is asymptotically optimal with respect to the number of samples under the assumption that the  subposteriors are Gaussian. The second idea we build on is over-the-air computing in order to enhance spectral efficiency. This is in line with previous works on federated (frequentist) learning in wireless systems \cite{zhu2019broadband,amiri2020machine,yang2020federated,liu2020privacy}.  although the goal here is synthesizing samples from the posterior and not computing an average.  
 For arbitrary subposteriors, building on these two ideas, we have proposed a general variational strategy termed wireless variational CMC (WVCMC) that aims at minimizing the KL divergence between the distributions of the output of the aggregation function at the server and of the global posterior.  Simulation results show that the proposed approach can vastly outperform existing solutions.  
 
{\color{black}We note that  performance of the proposed WCMC schemes depends on the  aggregation function. Although only linear mappings are considered in this paper in line with prior work on the topic, the introduced variational framework can be directly applied to other functions. }
{\color{black}As  for future work, one may consider studying channel-driven sampling for digital transmission via lattice coding \cite{zamir2014lattice}.} It would also be interesting to alleviate the assumption that workers can generate samples from local subposteriors. A possible approach is to carry out approximate MC sampling at workers, e.g., distributed SGLD \cite{ahn2014distributed} or distributed SVGD \cite{kassab2020federated}.   As another possible extension of the current work, optimal analog encoding and analog compression in \cite{abdi2020analog, abdi2020quantized}, could be  tailored for wireless CMC strategies so as to further enhance the communication efficiency. {\color{black} One could also enforce a short-term power constraint in each communication block. A possible way to do this involves introducing a power control threshold, in a manner similar to \cite{cao2020optimized,liu2020privacy}. Accordingly, workers whose channel gain is too small refrain from transmitting. The optimization of the thresholds is an interesting open problem.}

\appendix 

\subsection{Proof of Lemma~\ref{lemma: LB entropy}} \label{proof entropy}
According to aggregated function  in \eqref{eq: est theta} and received signal in \eqref{eq: rev signal},  we have 
\begin{align}
&{\sf H}[ q_F(\bm \theta)] \nn\\
&={\sf H}\Big[ \sum_{k=1}^K  \bW_k \by_k\Big]\nn\\
&={\sf H}\Big[ \sum_{k=1}^K  \bW_k ( \bE_k {\bm \theta}_k+  {{\bn}_k}) \Big]  \nn\\
&\overset{(a)}{\geq} \frac{d}{2}\log \Big[\sum_{k=1}^K  e^{\frac{2}{d} {\sf H}[\bW_k\bE_k {\bm \theta}_k]}+e^{\frac{2}{d} {\sf H}[\bW_k \bn_k]}\Big] \nn
\end{align}
\begin{align}
&\overset{(b)}{\geq}\frac{d}{2}\log \Big[ 2Ke^{\frac{1}{2K}\sum_{k=1}^K \frac{2}{d} {\sf H}[\bW_k\bE_k {\bm \theta}_k] +\sum_{k=1}^K\frac{2}{d} {\sf H}[\bW_k \bn_k]}\Big]  \nn\\
&=\frac{d}{2}\log(2K)+ \frac{1}{2K}\bigg(\sum_{k=1}^K {\sf H}[\bW_k\bE_k {\bm \theta}_k] + {\sf H}[\bW_k \bn_k] \bigg)\nn\\
&=\frac{d}{2}\log(2K)+ \frac{1}{2K}\bigg(\sum_{k=1}^K \log |\det (\bW_k\bE_k)|+{\sf H}[p_k({\bm \theta}_k|\cZ_k)] \nn\\ 
&\qquad  \qquad \qquad \qquad \qquad \qquad + \frac{1}{2} \log \det{2\pi e N_o \bW_k\bW_k^{\sf T}}\bigg) \nn 
\end{align}
where $(a)$ is obtained by using entropy power inequality, and $(b)$ follows  from  Jensen's inequality for the convex exponential function.  Rearranging terms in the result  yields   Lemma~\ref{lemma: LB entropy}.

 \subsection{Implementation of Bayesian Probit Regression} \label{appendix: implementation}
 
 \subsubsection{Gradient Computation}We assume the prior  follows Gaussian distribution $\mathcal{N}(0, \sigma^2 \bI) $, and the log prior is 
 \begin{align}
 \log p({\bm \theta}) = -\frac{1}{2 \sigma^2 }\|{\bm \theta}\|^2
 \end{align}
 According to \eqref{eq: probit model}, we have log likelihood as 
\begin{align}
\log p(v_{n}|{\bm \theta},\bu_{n}) \!  =\!  v_{n} \log \Phi({\bm \theta}^{\sf T} \bu_{n})  \!+\!(\!1\!-\! v_{n}\!)\!\log\!\l(\!1\!- \! \Phi({\bm \theta}^{\sf T} \bu_{n})\r),
\end{align}
where $\Phi(\cdot)$ is cumulative distribution function of the standard normal distribution. Accordingly, their gradients are given as 
\begin{align}
\nabla_{{\bm \theta}}\log p({\bm \theta}) &= -\frac{1}{ \sigma^2 }{\bm \theta}, \label{eq: gradient theta prior probit}\\
\nabla_{{\bm \theta}} \log p(v_{n}|{\bm \theta},\bu_n)  & = \frac{\phi({\bm \theta}^{\sf T} \bu_{n})\l[v_{n}-\Phi({\bm \theta}^{\sf T} \bu_{n})\r]}{\Phi({\bm \theta}^{\sf T} \bu_{n})\l[1-\Phi({\bm \theta}^{\sf T} \bu_{n})\r]}\bu_{n} \label{eq: gradient theta posterior probit}
\end{align}
where $\phi(x)=\frac{1}{\sqrt{2\pi}}e^{-{x^2}/{2}}$.

 \subsubsection{Gibbs Sampling from Posterior}
The analytical approach to perform Bayesian inference on probit model \eqref{eq: probit model} is challenging due to the lack of conjugate prior for the parameter $\bm \theta$.  
 An equivalent model can be provided by introducing an auxiliary variable ${ \kappa}_n$ as 
 \begin{subequations}\label{eq: equivalent model}
\begin{align} 
{ \kappa_n}&={\bm \theta}^{\sf T} \bu_n+\epsilon\\
v_n &=\left\{\begin{matrix}
1,  {\text {if } } { \kappa_n}>0\\ 
0,  {\text {otherwise,}}
\end{matrix}\right.
\end{align}
 \end{subequations}
where $\epsilon\sim \mathcal{N}(0,1)$.  
The equivalent model in \eqref{eq: equivalent model} enables an efficient Gibbs sampling for obtaining the posterior statistics.  The Gibbs sampling, as introduced in \cite{albert1993bayesian},  involves iterative sampling of the following two posterior distributions of $\bm \theta$ and $\kappa_n$ respectively. 

For $N$ observations, given the latent variables $\{\kappa_n\}_{n=1}^{N}$ and the covariates $ \{\bu_n\}_{n=1}^N$,  the posterior distribution of  ${\bm \theta}$ is derived as 
\begin{align}
p({\bm \theta}|\{\kappa_n\}_{n=1}^N, \{\bu_n\}_{n=1}^N)&\propto p(\bm \theta)  \prod_{n=1}^N p(\kappa_n|\bm \theta, \bu_n) \\
&=  \mathcal{N}_{\bm \theta}(0,\sigma^2\bI) \prod_{n=1}^N  \mathcal{N}_{\kappa_n}({\bm \theta}^{\sf T} \bu_n,1) \\
&\propto \mathcal{N}_{\bm \theta}\bigg( \bC \sum_{n=1}^N \bu_n \kappa_n,\bC  \bigg) \label{eq: sample posterior}
\end{align}
where $\bC=\Big(\sum_{n=1}^N \bu_n\bu_n^{\sf T} +\sigma^{-2}\bI \Big)^{-1}$.

On the other hand, with the knowledge of $\bm \theta$, $\bu_n$  and $v_n$, the posterior distribution of $\kappa_n$ is truncated Gaussian distribution 
\begin{align}\label{eq: sample latent variable}
p(\kappa_n|{\bm \theta}, \bu_n ,v_n)\propto\left\{\begin{matrix}
\mathcal{N}_{\kappa_n}({\bm \theta}^{\sf T} \bu_n, 1)  \mathcal{I}(\kappa_n>0, v_n=1)\\ 
\mathcal{N}_{\kappa_n}({\bm \theta}^{\sf T} \bu_n, 1)  \mathcal{I}(\kappa_n\leq0, v_n=0)\end{matrix},
\right.
\end{align}
where ${\mathcal{I}}(\cdot)$ is the indicator function. 

To start the sampling, the initial value of $\bm \theta^{(0)}$ can be set as the maximum likelihood estimate.   The samples of (sub)posterior are drawn from \eqref{eq: sample posterior} while iterating between \eqref{eq: sample posterior}  and \eqref{eq: sample latent variable}. The samples of global posterior $\{{\bm \theta}_g^{(s)}\}_{s=1}^{S_g}$  are obtained by using global data set and prior $\cN(0, \sigma^2 \bI)$, and $S_g=20000$, while the samples of  subposterior  $\{{\bm \theta}_k^{(s)}\}_{s=1}^{S}$ are based on the local data set and prior $\cN(0, K\sigma^2 \bI)$. The first $100$ samples are considered  to be burn-in phase, which are excluded in the evaluation. 

{\color{black}
\subsection{On the Complexity of Sampling from the Posterior}\label{app: sampling complexity}

The complexity of drawing one or multiple samples from the posterior $p({\bm \theta}|\mathcal{Z})$ for some data set $\mathcal{Z}$, prior $p({\bm \theta})$ and likelihood $p(\mathcal{Z}| {\bm \theta})$ depends strongly on the choice of prior and likelihood. At one extreme, if prior and likelihood are conjugate, the per-sample complexity is constant, and it does not depend on the size of the data set $\mathcal{Z}.$ At the other extreme, we have general-purpose MCMC algorithms such as rejection sampling, Metropolis-Hastings, or Langevin Monte Carlo that apply to any prior $p({\bm \theta})$ and likelihood $p(\mathcal{Z}|{\bm \theta})$. In between, we have solutions like Gibbs sampling -- a special instance of Metropolis-Hastings algorithms -- that is only relevant for priors and likelihoods that have suitable factorizations.

The problem of quantifying the complexity of general-purpose MCMC schemes is generally open \cite{bhatnagar2011computational}, and recent work has provided bounds on iteration complexity for specific classes of schemes such as SG-MCMC \cite{raginsky2017non,dalalyan2019user}. It is noted that these are approximate MC sampling schemes that produce samples from the posterior $p({\bm \theta}|\mathcal{Z})$ only asymptotically as the number of iterations grows. Methods such as SG-MCMC are justified by the fact that the per-iteration complexity scales linearly with the size of the mini-batch.

While, as mentioned, the general problem of relating accuracy, complexity, and number of data points is open, some insight on this issue can be obtained by considering the simple rejection sampling scheme. Rejection sampling produces exact samples from the posterior $p({\bm \theta}|\mathcal{Z})$ by first drawing samples from a proposal distribution, and then accepting each candidate sample with some probability. The complexity of rejection sampling can hence be directly measured by the average number of candidate samples needed to produce one sample from $p({\bm \theta}|\mathcal{Z})$.

A standard choice for the acceptance probability of a sample ${\bm \theta}$ is
\[
p(\mathcal{\textrm{acc}}|{\bm \theta})=\frac{p(\mathcal{Z}|{\bm \theta})}{p(\mathcal{Z}|{\bm \theta}_{\mathcal{}}^{ML})}=\prod_{n=1}^{N}\frac{p(\bz_{n}|{\bm \theta})}{p(\bz_{n}|{\bm \theta}_{\mathcal{}}^{ML})},
\]
where ${\bm \theta}_{\mathcal{}}^{ML}$ is the maximum likelihood (ML) solution $\ensuremath{{\bm \theta}_{\mathcal{}}^{ML}=\arg\max_{{\bm \theta}}p(\mathcal{Z}|{\bm \theta})}$.
By definition of the ML solution, the ratio ${p(\bz_{n}|{\bm \theta})}/{p(\bz_{n}|{\bm \theta}_{\mathcal{}}^{ML})}$ is generally smaller than one for a non-negligible fraction of data points, showing that the probability of acceptance can potentially decrease exponentially quickly to zero as $N$ increases. In turn, this implies that the complexity of generating a sample can grow as quickly as exponentially with the number of data points $N$. 
}
 
\bibliographystyle{ieeetr}

\end{document}

%% file: header.tex
\newtheorem{theorem}{Theorem}
\newtheorem{acknowledgement}[theorem]{Acknowledgement}
\newtheorem{axiom}[theorem]{Axiom}
\newtheorem{case}[theorem]{Case}
\newtheorem{claim}[theorem]{Claim}
\newtheorem{conclusion}[theorem]{Conclusion}
\newtheorem{condition}[theorem]{Condition}
\newtheorem{conjecture}[theorem]{Conjecture}
\newtheorem{criterion}[theorem]{Criterion}
\newtheorem{definition}{Definition}
\newtheorem{exercise}[theorem]{Exercise}
\newtheorem{lemma}{Lemma}
\newtheorem{corollary}{Corollary}
\newtheorem{notation}[theorem]{Notation}
\newtheorem{problem}[theorem]{Problem}
\newtheorem{proposition}{Proposition}
\newtheorem{scheme}{Scheme}   
\newtheorem{solution}[theorem]{Solution}
\newtheorem{summary}[theorem]{Summary}
\newtheorem{assumption}{Assumption}
\newtheorem{example}{\bf Example}
\newtheorem{remark}{\bf Remark}

\def\qed{$\Box$}
\def\QED{\mbox{\phantom{m}}\nolinebreak\hfill$\,\Box$}
\def\proof{\noindent{\emph{Proof:} }}
\def\poof{\noindent{\emph{Sketch of Proof:} }}
\def
\endproof{\hspace*{\fill}~\qed
\par
\endtrivlist\unskip}
\def\endproof{\hspace*{\fill}~\qed\par\endtrivlist\vskip3pt}

\def\E{\mathsf{E}}
\def\eps{\varepsilon}
\def\phi{\varphi}
\def\Lsp{{\boldsymbol L}}
\def\Bsp{{\boldsymbol B}}
\def\lsp{{\boldsymbol\ell}}
\def\Ltsp{{\Lsp^2}}
\def\Lpsp{{\Lsp^p}}
\def\Linsp{{\Lsp^{\infty}}}
\def\LtR{{\Lsp^2(\Rst)}}
\def\ltZ{{\lsp^2(\Zst)}}
\def\ltsp{{\lsp^2}}
\def\ltZt{{\lsp^2(\Zst^{2})}}
\def\ninN{{n{\in}\Nst}}
\def\oh{{\frac{1}{2}}}
\def\grass{{\cal G}}
\def\ord{{\cal O}}
\def\dist{{d_G}}
\def\conj#1{{\overline#1}}
\def\ntoinf{{n \rightarrow \infty }}
\def\toinf{{\rightarrow \infty }}
\def\tozero{{\rightarrow 0 }}
\def\trace{{\operatorname{Tr}}}
\def\ord{{\cal O}}
\def\UU{{\cal U}}
\def\rank{{\operatorname{rank}}}
\def\acos{{\operatorname{acos}}}

\def\SINR{\mathsf{SINR}}
\def\SNR{\mathsf{SNR}}
\def\SIR{\mathsf{SIR}}
\def\tSIR{\widetilde{\mathsf{SIR}}}
\def\Ei{\mathsf{Ei}}
\def\l{\left}
\def\r{\right}
\def\({\left(}
\def\){\right)}
\def\lb{\left\{}
\def\rb{\right\}}

\setcounter{page}{1}

\newcommand{\eref}[1]{(\ref{#1})}
\newcommand{\fig}[1]{Fig.\ \ref{#1}}

\def\bydef{:=}
\def\ba{{\mathbf{a}}}
\def\bb{{\mathbf{b}}}
\def\bc{{\mathbf{c}}}
\def\bd{{\mathbf{d}}}
\def\bee{{\mathbf{e}}}
\def\bff{{\mathbf{f}}}
\def\bg{{\mathbf{g}}}
\def\bh{{\mathbf{h}}}
\def\bi{{\mathbf{i}}}
\def\bj{{\mathbf{j}}}
\def\bk{{\mathbf{k}}}
\def\bl{{\mathbf{l}}}
\def\bn{{\mathbf{n}}}
\def\bo{{\mathbf{o}}}
\def\bp{{\mathbf{p}}}
\def\bq{{\mathbf{q}}}
\def\br{{\mathbf{r}}}
\def\bs{{\mathbf{s}}}
\def\bt{{\mathbf{t}}}
\def\bu{{\mathbf{u}}}
\def\bv{{\mathbf{v}}}
\def\bw{{\mathbf{w}}}
\def\bx{{\mathbf{x}}}
\def\by{{\mathbf{y}}}
\def\bz{{\mathbf{z}}}
\def\b0{{\mathbf{0}}}

\def\bA{{\mathbf{A}}}
\def\bB{{\mathbf{B}}}
\def\bC{{\mathbf{C}}}
\def\bD{{\mathbf{D}}}
\def\bE{{\mathbf{E}}}
\def\bF{{\mathbf{F}}}
\def\bG{{\mathbf{G}}}
\def\bH{{\mathbf{H}}}
\def\bI{{\mathbf{I}}}
\def\bJ{{\mathbf{J}}}
\def\bK{{\mathbf{K}}}
\def\bL{{\mathbf{L}}}
\def\bM{{\mathbf{M}}}
\def\bN{{\mathbf{N}}}
\def\bO{{\mathbf{O}}}
\def\bP{{\mathbf{P}}}
\def\bQ{{\mathbf{Q}}}
\def\bR{{\mathbf{R}}}
\def\bS{{\mathbf{S}}}
\def\bT{{\mathbf{T}}}
\def\bU{{\mathbf{U}}}
\def\bV{{\mathbf{V}}}
\def\bW{{\mathbf{W}}}
\def\bX{{\mathbf{X}}}
\def\bY{{\mathbf{Y}}}
\def\bZ{{\mathbf{Z}}}

\def\mA{{\mathbb{A}}}
\def\mB{{\mathbb{B}}}
\def\mC{{\mathbb{C}}}
\def\mD{{\mathbb{D}}}
\def\mE{{\mathbb{E}}}
\def\mF{{\mathbb{F}}}
\def\mG{{\mathbb{G}}}
\def\mH{{\mathbb{H}}}
\def\mI{{\mathbb{I}}}
\def\mJ{{\mathbb{J}}}
\def\mK{{\mathbb{K}}}
\def\mL{{\mathbb{L}}}
\def\mM{{\mathbb{M}}}
\def\mN{{\mathbb{N}}}
\def\mO{{\mathbb{O}}}
\def\mP{{\mathbb{P}}}
\def\mQ{{\mathbb{Q}}}
\def\mR{{\mathbb{R}}}
\def\mS{{\mathbb{S}}}
\def\mT{{\mathbb{T}}}
\def\mU{{\mathbb{U}}}
\def\mV{{\mathbb{V}}}
\def\mW{{\mathbb{W}}}
\def\mX{{\mathbb{X}}}
\def\mY{{\mathbb{Y}}}
\def\mZ{{\mathbb{Z}}}

\def\cA{\mathcal{A}}
\def\cB{\mathcal{B}}
\def\cC{\mathcal{C}}
\def\cD{\mathcal{D}}
\def\cE{\mathcal{E}}
\def\cF{\mathcal{F}}
\def\cG{\mathcal{G}}
\def\cH{\mathcal{H}}
\def\cI{\mathcal{I}}
\def\cJ{\mathcal{J}}
\def\cK{\mathcal{K}}
\def\cL{\mathcal{L}}
\def\cM{\mathcal{M}}
\def\cN{\mathcal{N}}
\def\cO{\mathcal{O}}
\def\cP{\mathcal{P}}
\def\cQ{\mathcal{Q}}
\def\cR{\mathcal{R}}
\def\cS{\mathcal{S}}
\def\cT{\mathcal{T}}
\def\cU{\mathcal{U}}
\def\cV{\mathcal{V}}
\def\cW{\mathcal{W}}
\def\cX{\mathcal{X}}
\def\cY{\mathcal{Y}}
\def\cZ{\mathcal{Z}}
\def\cd{\mathcal{d}}
\def\Mt{M_{t}}
\def\Mr{M_{r}}
\def\O{\Omega_{M_{t}}}
\newcommand{\figref}[1]{{Fig.}~\ref{#1}}
\newcommand{\tabref}[1]{{Table}~\ref{#1}}

\newcommand{\var}{\mathsf{var}}
\newcommand{\fb}{\tx{fb}}
\newcommand{\nf}{\tx{nf}}
\newcommand{\BC}{\tx{(bc)}}
\newcommand{\MAC}{\tx{(mac)}}
\newcommand{\Pout}{p_{\mathsf{out}}}
\newcommand{\nnn}{\nn\\}
\newcommand{\FB}{\tx{FB}}
\newcommand{\TX}{\tx{TX}}
\newcommand{\RX}{\tx{RX}}
\renewcommand{\mod}{\tx{mod}}
\newcommand{\m}[1]{\mathbf{#1}}
\newcommand{\td}[1]{\tilde{#1}}
\newcommand{\sbf}[1]{\scriptsize{\textbf{#1}}}
\newcommand{\stxt}[1]{\scriptsize{\textrm{#1}}}
\newcommand{\suml}[2]{\sum\limits_{#1}^{#2}}
\newcommand{\sumlk}{\sum\limits_{k=0}^{K-1}}
\newcommand{\eqhsp}{\hspace{10 pt}}
\newcommand{\tx}[1]{\texttt{#1}}
\newcommand{\Hz}{\ \tx{Hz}}
\newcommand{\sinc}{\tx{sinc}}
\newcommand{\tr}{\mathrm{tr}}
\newcommand{\diag}{\mathrm{diag}}
\newcommand{\MAI}{\tx{MAI}}
\newcommand{\ISI}{\tx{ISI}}
\newcommand{\IBI}{\tx{IBI}}
\newcommand{\CN}{\tx{CN}}
\newcommand{\CP}{\tx{CP}}
\newcommand{\ZP}{\tx{ZP}}
\newcommand{\ZF}{\tx{ZF}}
\newcommand{\SP}{\tx{SP}}
\newcommand{\MMSE}{\tx{MMSE}}
\newcommand{\MINF}{\tx{MINF}}
\newcommand{\RC}{\tx{MP}}
\newcommand{\MBER}{\tx{MBER}}
\newcommand{\MSNR}{\tx{MSNR}}
\newcommand{\MCAP}{\tx{MCAP}}
\newcommand{\vol}{\tx{vol}}
\newcommand{\ah}{\hat{g}}
\newcommand{\tg}{\tilde{g}}
\newcommand{\teta}{\tilde{\eta}}
\newcommand{\heta}{\hat{\eta}}
\newcommand{\uh}{\m{\hat{s}}}
\newcommand{\eh}{\m{\hat{\eta}}}
\newcommand{\hv}{\m{h}}
\newcommand{\hh}{\m{\hat{h}}}
\newcommand{\Po}{P_{\mathrm{out}}}
\newcommand{\Poh}{\hat{P}_{\mathrm{out}}}
\newcommand{\Ph}{\hat{\gamma}}
\newcommand{\mat}[1]{\begin{matrix}#1\end{matrix}}
\newcommand{\ud}{^{\dagger}}
\newcommand{\C}{\mathcal{C}}
\newcommand{\nn}{\nonumber}
\newcommand{\nInf}{U\rightarrow \infty}